%% file: main.tex
\documentclass[conference]{IEEEtran}
\pdfoutput=1

\usepackage[T1]{fontenc}
\usepackage[utf8]{inputenc}
\usepackage{cite}
\usepackage{amsmath,amssymb}
\usepackage[scaled=0.83]{beramono}
\usepackage{booktabs}
\usepackage{tabularx}
\usepackage{array}
\usepackage{listings}
\usepackage{tikz}
\usetikzlibrary{arrows.meta,positioning,calc}
\usepackage{pgfplots}
\pgfplotsset{compat=1.18}
\usepackage[hyphens]{url}
\usepackage[hidelinks]{hyperref}
\usepackage{orcidlink}
\usepackage{microtype}

\let\origunderscore\_
\newcommand{\ubrk}{\origunderscore\allowbreak}
\DeclareRobustCommand{\code}[1]{\texorpdfstring{\ifmmode\texttt{#1}\else{\begingroup\let\_\ubrk\fontshape{n}\ttfamily #1\endgroup}\fi}{#1}}
\input{generated/numbers}

\begin{document}

\title{Power-Aware Simulation and Formal Equivalence for Unmodified SystemC and SystemVerilog RTL}

\author{\IEEEauthorblockN{Mayur Kubavat\,\orcidlink{0009-0009-1087-8558}}
\IEEEauthorblockA{\textit{Independent Researcher, Mountain View, CA, USA}}}

\maketitle
\thispagestyle{plain}

\begin{abstract}
Modern systems-on-chip specify power gating, domain isolation, and state retention in external IEEE Std 1801 Unified Power Format (\code{.upf}) Tcl scripts so that RTL modules stay portable across simulation, synthesis, and formal sign-off.
C++17 verification environments built on SystemC, Verilator, and UVM-SystemC have no runtime for those scripts: prior SystemC power libraries either estimate transaction-level energy without corrupting hardware state or embed custom C++ macros inside RTL modules, breaking synthesis and formal equivalence. Forcing signals from external observers also violates IEEE~1666-2023 scheduler and single-writer invariants, while synthesizable SystemC integers and Verilator models use two-state types that drop Verilog \code{'X} values and mask missing isolation cells.
We present \code{systemc-lp}, a C++17 IEEE~1801 engine built on a UPF-agnostic, \code{+\SCHOOKDIFF{}}-line policy interface in the SystemC~3.0.3 kernel. Keeping power intent out of RTL lets one unmodified design and \code{.upf} specification drive bypassed functional runs when UPF is disabled; cycle-accurate simulation with process gating, four-state \code{'X}/\code{'Z} and two-state \code{CORRUPT\_JITTER} corruption, and header-only UVM/TLM/FC4SC adapters when active; and word-level Btor2 and SystemVerilog formal proofs in \code{svcc} and \code{InFormal}.
Across the open Accellera C++17 verification stack (\code{\SCCTESTS{}/\SCCTESTS{}} SystemC kernel tests, \code{ASan+UBSan} clean) and an APB PMU-timer subsystem, disabling UPF stays within \LPOFFDELTA{} of unhooked SystemC, formal proofs finish within 70\,ms, and cross-checking all three tools uncovered three bugs in \code{systemc-lp} alongside six RTL, UPF, and firmware-sequencing defects missed by directed simulation.
\end{abstract}

\begin{IEEEkeywords}
low-power verification, IEEE 1801 UPF, power-aware simulation, SystemC, Verilator, UVM-SystemC, formal equivalence checking, self-composition.
\end{IEEEkeywords}

\input{sections/introduction}
\input{sections/overview}
\input{sections/architecture}
\input{sections/semantics_and_dv}
\input{sections/formal_continuity}
\input{sections/evaluation}
\input{sections/related_work}
\input{sections/conclusion}

\IEEEtriggeratref{25}
\bibliographystyle{IEEEtran}
\bibliography{refs}

\end{document}

%% file: generated/numbers.tex
\newcommand{\SCLPCOMMIT}{3739b0d}
\newcommand{\SCCOMMIT}{bf26f38}
\newcommand{\UVMCOMMIT}{76dee61}
\newcommand{\CRAVECOMMIT}{3f42ff7}
\newcommand{\FCFOURCOMMIT}{adb746d}
\newcommand{\EXCOMMIT}{de370d7}
\newcommand{\SVCCCOMMIT}{934cf85}
\newcommand{\INFCOMMIT}{99f6f99}
\newcommand{\INFCRATES}{29}

\newcommand{\SCHOOKDIFF}{256}
\newcommand{\SCHOOKLOC}{378}

\newcommand{\SCLPTOTALLOC}{12,164}
\newcommand{\SCLPSRCLOC}{7,958}

\newcommand{\SCLPCTESTS}{30}
\newcommand{\SCLPTIME}{0.44\,s}
\newcommand{\APBSUBSYSLOC}{949}

\newcommand{\SVCCUPFADD}{5,750}

\newcommand{\SCCTESTS}{969}
\newcommand{\SCEXAMPLES}{35}

\newcommand{\UVMCTESTS}{177}
\newcommand{\CRAVECTESTS}{530}
\newcommand{\FCFOURCTESTS}{290}
\newcommand{\SMOKEPASS}{11}

\newcommand{\LPOFFDELTA}{5\%}

%% file: sections/introduction.tex
\section{Introduction}
\label{sec:intro}

Modern systems-on-chip (SoCs) partition logic into always-on (\code{AON}), switchable, state-retaining, and voltage-scaled power domains controlled by an on-chip Power Management Unit (PMU). To keep register-transfer level (RTL) descriptions portable across simulation, logic synthesis, and formal sign-off, engineers write power domains, supply switches, isolation clamps, retention registers, and Power State Tables (PSTs) in external IEEE Std~1801 Unified Power Format (\code{.upf}) Tcl scripts~\cite{ieee1801_2018,ieee1801_2024}. Over the same period, hardware verification teams have increasingly compiled SystemVerilog and synthesizable SystemC~\cite{ieee1666_1_2016} into C++17 executables using Verilator~\cite{snyder_verilator} and SystemC~\cite{ieee1666_2023}. Paired with Accellera's UVM-SystemC~\cite{ieee1800_2_2020}, CRAVE constrained-random stimulus, and FC4SC functional coverage with Unified Coverage Interoperability Standard (UCIS~1.0)~\cite{accellera_ucis} reporting, compiled C++17 testbenches run at megacycle-per-second host speeds under standard memory and undefined-behavior sanitizers while following IEEE~1800-2023~\cite{ieee1800_2023} verification methodologies.

\input{figures/fig_arch}

That C++17 verification stack stops short at low-power sign-off. Commercial four-state Verilog simulators elaborate \code{.upf} scripts inside proprietary kernels to clamp isolated domain crossings, snapshot retention registers prior to shutdown, and propagate unknown (\code{'X}) values across powered-down logic~\cite{bembaron2009dvcon}. Open C++17 environments have no equivalent runtime. Earlier SystemC power tools either attach passive energy estimators that leave architectural state unchanged~\cite{vece2009pktool,vece2011pktool}, model coarse transaction-level sockets above clock edges and pins~\cite{lebreton2008pwarch,mbarek2012upftlm,mischkalla2013patmos}, or require designers to rewrite RTL registers and modules with custom C++ macros~\cite{gagarski2016scpower}. Intrusive source macros break synthesizable SystemC compilers such as \code{svcc}~\cite{kubavat2026svcc} and prevent formal engines~\cite{hazra2013powertructor,kubavat2026informal} from verifying the exact files exercised in simulation. Executing external \code{.upf} specifications without editing RTL raises three technical problems. First, production \code{.upf} files are executable Tcl programs with loops and wildcard hierarchy queries rather than static tables. Second, forcing signals from an external SystemC observer leaves off-domain processes running on every clock edge, trips single-writer checks (\code{SC\_ONE\_WRITER}), and triggers IEEE~1666-2023 stage-callback warning \code{W553}. Third, although SystemC provides four-state \code{sc\_logic}/\code{sc\_lv<W>} for pads and tri-state buses, \code{sc\_lv<W>} defines no arithmetic operators in the IEEE~1666.1 synthesizable subset and \code{Verilator {-}{-}sc} lowers SystemVerilog \code{logic} to two-state C++ integers (\code{bool}, \code{uint32\_t}, \code{sc\_uint<W>}); without \code{'X}, zeroing or inverting powered-down words masks missing isolation cells whenever the corrupted bit matches a safe idle level.

Our central observation is that a compact, UPF-agnostic policy interface inside the SystemC~3.0.3 kernel (\code{+\SCHOOKDIFF{}} diff lines across modules, processes, and signal channels, with zero Tcl dependencies) is sufficient to execute IEEE~1801 UPF~3.1 semantics externally while leaving synthesizable C++ and SystemVerilog RTL unmodified (Fig.~\ref{fig:arch}). Built on those three hooks, \code{systemc-lp} embeds \code{libtcl} to evaluate 28 IEEE~1801 RTL commands plus two sign-off extensions (\code{add\_pst\_transition} and \code{set\_inrush\_limit}), suspends off-domain \code{SC\_METHOD} and \code{SC\_CTHREAD} dispatch, drives \code{SC\_LOGIC\_X}/\code{SC\_LOGIC\_Z} on four-state nets, and applies multi-cycle \code{CORRUPT\_JITTER} scrambling on two-state nets so sticky single-bit X-masking decays as $2^{-K}$ across $K$ off-domain clock edges. A CMake post-processing step binds flattened \code{Verilator {-}{-}sc} flip-flops and dual-rail SRAM arrays onto the same supply graph, while header-only adapters synchronize UVM-SystemC register mirrors, TLM-2.0 Direct Memory Interface (DMI) pointers, and FC4SC assertions without modifying those libraries for low-power support. Because the RTL carries no power macros, \code{svcc} and \code{InFormal} read the identical RTL and \code{.upf} files to discharge word-level Btor2~\cite{niemetz2018btor2} two-copy self-composition miters~\cite{yang2018lazyself,tan2026secic3} in Z3~\cite{demoura2008z3} alongside inductive SystemVerilog Assertions (SVA).

We validated the complete C++17 and Rust flow under Release and \code{ASan+UBSan} gates, passing all \code{\SCCTESTS{}/\SCCTESTS{}} SystemC~3.0.3 kernel tests and \code{\UVMCTESTS{}}, \code{\CRAVECTESTS{}}, and \code{\FCFOURCTESTS{}} tests in UVM-SystemC, CRAVE, and FC4SC. On an APB3 PMU-timer subsystem, disabling UPF adds at most \LPOFFDELTA{} runtime overhead relative to unhooked SystemC~3.0.3, formal equivalence and property proofs finish within 70\,ms, and running all three tools on the same design files exposed three latent bugs in \code{systemc-lp} together with six RTL, UPF, and firmware-sequencing defects that escaped directed tests. Section~\ref{sec:overview} introduces IEEE~1801 semantics on a concrete excerpt of that APB subsystem. Section~\ref{sec:arch} details the SystemC kernel hooks, delta-cycle schedule, and Tcl elaborator; Section~\ref{sec:semantics} covers four-state and two-state power-collapse semantics, Verilator binding, and testbench adapters; and Section~\ref{sec:formal} presents word-level formal lowering and dual-path model checking. Section~\ref{sec:eval} reports empirical overhead, fault-injection, formal mutation, and cross-tool bug-discovery results, followed by related work in Section~\ref{sec:related} and conclusions in Section~\ref{sec:conclusion}.

%% file: figures/fig_arch.tex
\begin{figure*}[t]
\centering
\definecolor{aonblue}{RGB}{28,69,135}
\definecolor{timrust}{RGB}{175,65,25}
\definecolor{upfteal}{RGB}{0,115,115}
\definecolor{hazred}{RGB}{175,30,30}
\begin{tikzpicture}[
  x=1mm, y=1mm,
  font=\sffamily\scriptsize,
  >=Stealth,
  modbox/.style={draw=black!80, line width=0.55pt, fill=white, rounded corners=1pt},
  regbox/.style={draw=black!85, line width=0.5pt, fill=white, minimum height=6.5mm, inner xsep=2.8pt, align=center},
  upfbox/.style={draw=upfteal, dashed, line width=0.65pt, fill=upfteal!7, rounded corners=1pt, align=center},
  engbox/.style={draw=black!80, line width=0.5pt, fill=black!3, rounded corners=1pt, align=left},
  sig/.style={->, draw=black!80, line width=0.48pt},
  upfsig/.style={->, draw=upfteal, densely dashed, line width=0.55pt},
  rail/.style={line width=1.1pt},
  dot/.style={circle, fill=black!85, inner sep=0pt, minimum size=1.1mm}
]

  \node[anchor=west, font=\sffamily\footnotesize\bfseries] at (0.5,69.4) {(a) Unmodified \code{apb\_pmu\_timer\_subsys} RTL with External \code{apb\_subsys.upf} Overlay};

  \draw[draw=upfteal, dashed, line width=0.65pt, fill=upfteal!7, rounded corners=1pt] (52.2,59.4) rectangle (64.8,64.4);
  \node[anchor=south, text=upfteal, font=\sffamily\tiny\bfseries] at (58.5,64.5) {SW\_TIM};
  \draw[rail, draw=aonblue] (9.5,61.8) -- (55.4,61.8);
  \draw[rail, draw=timrust, densely dashed] (61.6,61.8) -- (109.0,61.8);
  \draw[draw=upfteal, line width=0.95pt] (55.9,62.0) -- (61.1,63.6);
  \draw[draw=upfteal, fill=white, line width=0.6pt] (55.4,61.8) circle (0.55mm);
  \draw[draw=upfteal, fill=white, line width=0.6pt] (61.6,61.8) circle (0.55mm);

  \node[anchor=south west, text=aonblue, font=\sffamily\tiny\bfseries] at (9.5,62.1) {VDD\_AON (0.90V / 0.75V)};
  \draw[draw=aonblue, line width=0.75pt] (29.5,61.8) -- (29.5,56.5);
  \fill[aonblue] (29.5,61.8) circle (0.55mm);

  \node[anchor=south east, text=timrust, font=\sffamily\tiny\bfseries] at (109.0,62.1) {VDD\_TIM (0.90V / 0.75V / OFF)};
  \draw[draw=timrust, densely dashed, line width=0.75pt] (82.0,61.8) -- (82.0,56.5);
  \fill[timrust] (82.0,61.8) circle (0.55mm);

  \draw[draw=aonblue, line width=0.75pt, fill=aonblue!3, rounded corners=1.5pt] (9.5,7.5) rectangle (49.5,56.5);
  \node[anchor=north west, text=aonblue, font=\sffamily\scriptsize\bfseries] at (10.5,56.0) {PD\_AON (Always-On)};

  \node[modbox, minimum width=36.0mm, minimum height=14.0mm] (upmu) at (29.5,44.8) {};
  \node[anchor=north west, font=\sffamily\scriptsize\bfseries] at (12.2,51.3) {\code{u\_pmu} {\normalfont\tiny(CSR \code{0x00..0x04})}};
  \node[anchor=west, font=\sffamily\tiny, align=left] at (12.2,42.8) {CSRs: \code{pwr\_en}, \code{iso\_en},\\[-0.4pt]\code{ret\_save/restore}, \code{por\_n}\\[-0.4pt]Status: \code{tim\_pwr\_ack}};

  \node[modbox, minimum width=36.0mm, minimum height=23.3mm] (fab) at (29.5,20.15) {};
  \node[anchor=north west, font=\sffamily\scriptsize\bfseries] at (12.2,31.3) {\code{fabric\_mux} {\normalfont\tiny(APB3 Firewall)}};

  \node[draw=black!75, fill=black!4, minimum width=22.0mm, minimum height=4.6mm, font=\sffamily\tiny] (andg) at (24.4,25.8) {\code{PSEL \& is\_tim \& !iso}};
  \draw[sig] (1.0,25.8) -- (andg.west) node[pos=0.38, above=-0.5pt, font=\sffamily\tiny\bfseries] {\code{PSEL}};

  \draw[draw=black!85, fill=white, line width=0.5pt] (31.2,13.0) -- (25.0,14.8) -- (25.0,19.2) -- (31.2,21.0) -- cycle;
  \node[font=\sffamily\tiny] at (29.8,19.2) {0};
  \node[font=\sffamily\tiny] at (29.8,14.8) {1};
  \node[font=\sffamily\tiny\bfseries] at (27.4,17.0) {MUX};

  \draw[sig] (36.8,19.2) node[right=-0.5pt, font=\sffamily\tiny] {\code{pmu\_rdata}} -- (31.2,19.2);
  \draw[sig] (28.1,10.4) -- (28.1,13.9);
  \node[anchor=west, font=\sffamily\tiny] at (28.5,11.0) {\code{is\_tim}};
  \draw[sig] (25.0,17.0) -- (1.0,17.0) node[pos=0.80, above=-0.5pt, font=\sffamily\tiny\bfseries] {\code{PRDATA}};

  \draw[draw=timrust, dashed, line width=0.8pt, fill=timrust!4, rounded corners=1.5pt] (63.5,7.5) rectangle (109.0,56.5);
  \node[anchor=north west, text=timrust, font=\sffamily\scriptsize\bfseries] at (64.5,56.0) {PD\_TIM: \code{u\_timer} (Switched)};

  \node[regbox, minimum width=19.8mm] (retdff) at (75.2,43.8) {\hspace*{2.0pt}\textbf{DFF} \code{cfg\_q, load\_q}\\[0pt]{\tiny\hspace*{2.0pt}no \code{por\_n} (\code{cfg\_q})}};
  \draw ([yshift=-0.9mm]retdff.west) -- ([xshift=1.3mm]retdff.west) -- ([yshift=0.9mm]retdff.west);

  \node[draw=upfteal, dashed, line width=0.7pt, fill=upfteal!9, rounded corners=3.5mm, minimum width=15.2mm, minimum height=7.0mm, font=\sffamily\tiny, text=upfteal, align=center] (shadow) at (100.3,43.8) {\textbf{RET\_TIM}\\[-0.6pt]{\tiny Balloon Latch}};
  \draw[->, draw=upfteal, line width=0.5pt] ([yshift=0.95mm]retdff.east) -- ([yshift=0.95mm]shadow.west) node[midway, above=-0.9pt, font=\sffamily\fontsize{4.5}{5}\selectfont, text=upfteal] {save};
  \draw[<-, draw=upfteal, line width=0.5pt] ([yshift=-0.95mm]retdff.east) -- ([yshift=-0.95mm]shadow.west) node[midway, below=-0.9pt, font=\sffamily\fontsize{4.5}{5}\selectfont, text=upfteal] {restore};

  \draw[draw=aonblue, line width=0.65pt] (shadow.north) -- (100.3,49.1);
  \draw[rail, draw=aonblue] (97.5,49.1) -- (103.1,49.1);
  \node[anchor=south, text=aonblue, font=\sffamily\fontsize{4.6}{5}\selectfont\bfseries] at (100.3,49.2) {SS\_AON};

  \node[draw=black!80, fill=white, rounded corners=1pt, minimum width=13.5mm, minimum height=4.6mm, font=\sffamily\tiny] (cmp) at (79.0,35.0) {\code{count==1}};

  \node[regbox, minimum width=26.2mm] (voldff) at (81.8,25.8) {\hspace*{2.2pt}\textbf{DFF} \code{count\_q, scratch\_q}\\[0pt]{\tiny\hspace*{2.2pt}volatile: scrambles when OFF}};
  \draw ([yshift=-0.9mm]voldff.west) -- ([xshift=1.3mm]voldff.west) -- ([yshift=0.9mm]voldff.west);

  \node[draw=black!80, fill=white, rounded corners=1pt, minimum width=16.0mm, minimum height=4.6mm, font=\sffamily\tiny] (rdm) at (85.0,14.8) {CSR Read};

  \draw[sig] (79.0,29.1) -- (cmp.south);
  \draw[sig] (85.0,22.5) -- (rdm.north);

  \draw[draw=upfteal, line width=0.68pt, fill=upfteal!10]
    (59.1,32.8) -- (56.6,32.8) arc (270:90:2.2) -- (59.1,37.2) -- cycle;
  \draw[draw=upfteal, fill=white, line width=0.55pt] (57.3,37.75) circle (0.52mm);
  \node[text=upfteal, font=\sffamily\tiny\bfseries] at (57.2,35.0) {\&};
  \node[anchor=north east, text=upfteal, font=\sffamily\fontsize{4.2}{5}\selectfont\bfseries, inner sep=0.5pt] at (57.0,32.5) {ISO\_TIM};

  \draw[draw=upfteal, line width=0.68pt, fill=upfteal!10]
    (59.1,12.6) -- (56.6,12.6) arc (270:90:2.2) -- (59.1,17.0) -- cycle;
  \draw[draw=upfteal, fill=white, line width=0.55pt] (57.3,17.55) circle (0.52mm);
  \node[text=upfteal, font=\sffamily\tiny\bfseries] at (57.2,14.8) {\&};
  \node[anchor=north, text=upfteal, font=\sffamily\fontsize{4.6}{5}\selectfont\bfseries] at (56.6,12.3) {ISO\_TIM (0)};

  \draw[sig] (cmp.west) -- (59.1,35.0);
  \node[above=-0.5pt, font=\sffamily\tiny] at (68.0,35.0) {\code{tim\_irq}};
  \draw[sig] (54.4,35.0) -- (1.0,35.0) node[pos=0.90, above=-0.5pt, font=\sffamily\tiny\bfseries] {\code{IRQ}};

  \draw[sig] (andg.east) -- (voldff.west);
  \node[above=-0.5pt, font=\sffamily\tiny] at (41.4,25.8) {\code{tim\_psel}};

  \draw[sig] (rdm.west) -- (59.1,14.8);
  \node[above=-0.5pt, font=\sffamily\tiny] at (70.2,14.8) {\code{tim\_rdata}};
  \draw[sig] (54.4,14.8) -- (31.2,14.8);

  \draw[<->, draw=upfteal, densely dashed, line width=0.55pt] (47.5,49.5) -| (58.5,59.4);
  \node[anchor=south west, text=upfteal, font=\sffamily\fontsize{4.4}{5}\selectfont] at (49.9,49.5) {pwr\_en/ack};
  \draw[upfsig] (47.5,39.3) -- (100.3,39.3) -- (shadow.south);
  \draw[upfsig] (47.5,38.3) -| (57.3,38.28);
  \draw[upfsig] (57.3,32.8) -- (57.3,18.08);

  \draw[draw=black!80, line width=0.55pt, fill=white] (9.5,3.8) rectangle (13.5,5.8);
  \node[anchor=west, font=\sffamily\tiny] at (14.2,4.8) {Unmodified C++/SV RTL (\code{0} LP macros)};
  \draw[draw=upfteal, line width=0.6pt, fill=upfteal!10] (60.6,3.8) -- (58.8,3.8) arc (270:90:1.0) -- (60.6,5.8) -- cycle;
  \node[anchor=west, font=\sffamily\tiny, text=upfteal] at (61.3,4.8) {External \code{apb\_subsys.upf} cells (switch, AND-clamp, balloon)};

  \node[anchor=west, font=\sffamily\footnotesize\bfseries] at (115.0,69.4) {(b) Shared-Source Simulation \& Formal Miter};

  \node[engbox, text width=60.0mm, inner sep=2.2pt] (b1) at (147.2,56.2) {
    {\bfseries 1. Functional Run (No \code{--upf} / \code{lp\_stub})}\\[1pt]
    {\tiny Per-timestep callbacks unregistered; \code{m\_power\_ext\_p == nullptr} fast path runs identical RTL binary.}
  };

  \node[engbox, draw=aonblue, fill=aonblue!4, text width=60.0mm, inner sep=2.2pt] (b2) at (147.2,41.4) {
    {\bfseries\color{aonblue} 2. \code{systemc-lp} Runtime + Testbench Adapters}\\[1pt]
    {\tiny $\bullet$ \code{--lp-lint} ($t=0$) $\cdot$ \code{sc\_process\_b} gate $\cdot$ \code{CORRUPT\_JITTER}\\[-0.3pt]
     $\bullet$ Header-only adapters: UVM RAL, TLM DMI, FC4SC SVA/UCIS}
  };

  \draw[draw=timrust, line width=0.55pt, fill=timrust!3, rounded corners=1pt] (115.8,4.0) rectangle (178.0,31.5);
  \node[anchor=north west, font=\sffamily\scriptsize\bfseries, text=timrust] at (116.8,31.0) {3. \code{svcc prove-upf} \& \code{InFormal} 2-Copy Miter};

  \node[draw=black!80, fill=white, minimum width=18.5mm, minimum height=6.4mm, font=\sffamily\tiny, align=center] (m0) at (138.6,21.6) {Copy $M^{(0)}$\\[-1pt]{\tiny corrupt $\xi^{(0)}$}};
  \node[draw=black!80, fill=white, minimum width=18.5mm, minimum height=6.4mm, font=\sffamily\tiny, align=center] (m1) at (138.6,11.2) {Copy $M^{(1)}$\\[-1pt]{\tiny corrupt $\xi^{(1)}$}};
  \node[font=\sffamily\tiny, align=right, anchor=east] (xin) at (125.0,16.4) {Shared\\[-0.5pt]inputs $X$};
  \draw[draw=black!80, line width=0.48pt] (xin.east) -- (126.8,16.4);
  \node[dot] at (126.8,16.4) {};
  \draw[sig] (126.8,16.4) |- (m0.west);
  \draw[sig] (126.8,16.4) |- (m1.west);

  \node[circle, draw=hazred, fill=white, line width=0.6pt, inner sep=1.2pt, font=\sffamily\scriptsize\bfseries, text=hazred] (xor) at (155.4,16.4) {$\neq$};
  \draw[sig] (m0.east) -| (xor.north);
  \draw[sig] (m1.east) -| (xor.south);
  \node[draw=black!80, fill=white, rounded corners=1pt, minimum width=13.6mm, minimum height=8.2mm, font=\sffamily\tiny, align=center] (z3) at (169.3,16.4) {\textbf{Z3 / IC3}\\[-0.5pt]2QBF check\\[-0.5pt]{\color{upfteal}\textbf{UNSAT}}};
  \draw[sig, draw=hazred] (xor.east) -- (z3.west);

  \draw[draw=black!65, line width=0.55pt] (109.0,32.0) -- (112.2,32.0);
  \draw[draw=black!65, line width=0.55pt] (112.2,17.8) -- (112.2,56.2);
  \node[dot] at (112.2,32.0) {};
  \node[dot] at (112.2,41.4) {};
  \draw[sig, line width=0.55pt] (112.2,56.2) -- (b1.west);
  \draw[sig, line width=0.55pt, draw=aonblue] (112.2,41.4) -- (b2.west);
  \draw[sig, line width=0.55pt, draw=timrust] (112.2,17.8) -- (115.8,17.8);

\end{tikzpicture}
\caption{Motivating architecture on \code{apb\_pmu\_timer\_subsys}. (a)~Synthesizable SystemC/SystemVerilog RTL contains no low-power macros; \code{apb\_subsys.upf} attaches power switch \code{SW\_TIM}, port-scoped \code{ISO\_TIM} clamps, and \code{RET\_TIM} shadow latches externally. (b)~The same RTL and \code{.upf} files drive bypassed functional runs, cycle-accurate \code{systemc-lp} simulation, and word-level two-copy self-composition proofs ($M^{(0)} \times M^{(1)}$) in \code{svcc} and \code{InFormal}.}
\label{fig:arch}
\end{figure*}
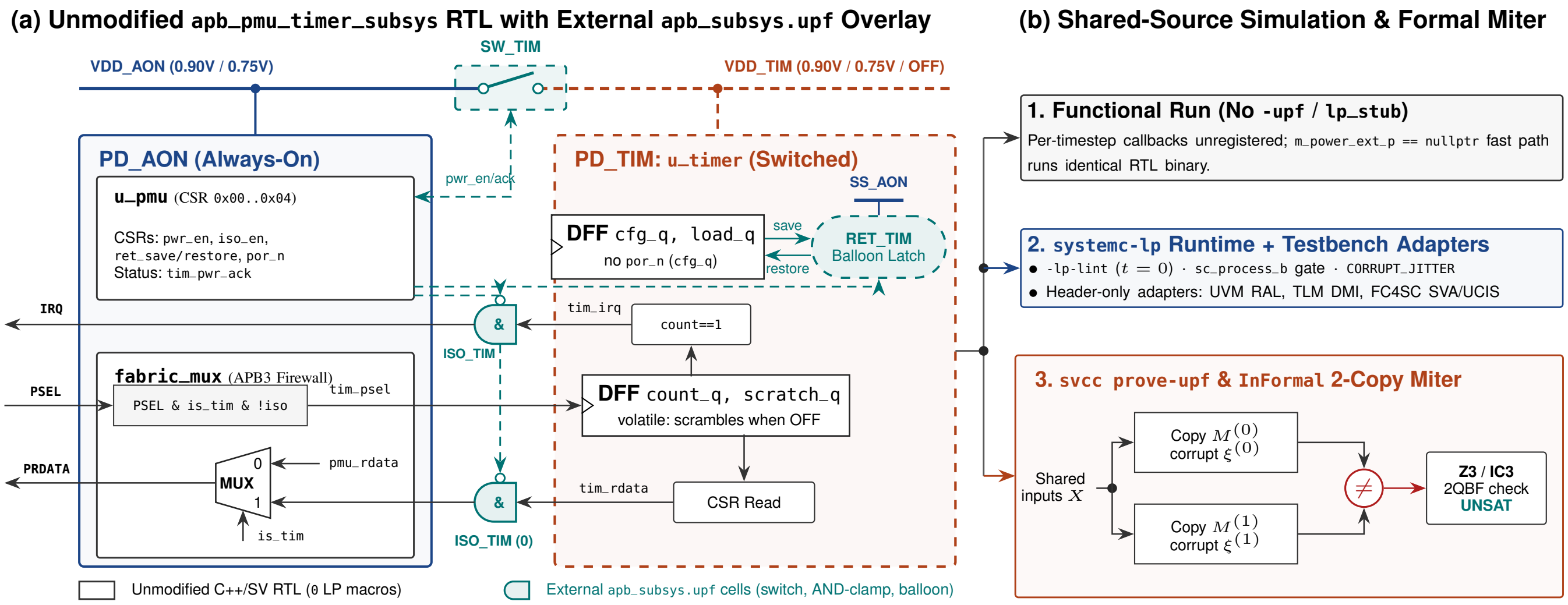

%% file: sections/overview.tex
\section{Overview and Motivating Example}
\label{sec:overview}

IEEE Std~1801~\cite{ieee1801_2018,ieee1801_2024} separates power architecture from functional hardware description. A standalone \code{.upf} script partitions an elaborated module tree into power domains (\code{create\_power\_domain}), each powered by a primary supply set (\code{create\_supply\_set}) pairing power and ground nets (\code{create\_supply\_net}). Supply ports and on-chip switches (\code{create\_power\_switch}) propagate discrete states in $\{\code{OFF}, \code{UNDETERMINED}, \code{PARTIAL\_ON}, \code{FULL\_ON}\}$ and voltage levels across the supply graph, mapping each domain onto one of four simulation states (\code{simstate}): normal execution (\code{NORMAL}), sequential and combinational loss (\code{CORRUPT}), sequential loss with held combinational outputs until inputs toggle (\code{CORRUPT\_ON\_ACTIVITY}), or drowsy standby that preserves state unless mutated (\code{CORRUPT\_STATE\_ON\_CHANGE}). Boundary isolation strategies (\code{set\_isolation}) clamp signals crossing from a collapsible source into an active sink, state-retention strategies (\code{set\_retention}) snapshot selected registers into always-on shadow latches across \code{save} and \code{restore} events, level shifters (\code{set\_level\_shifter}) guard cross-voltage nets, and Power State Tables (\code{create\_pst}) define legal multi-rail operating modes.

\begin{figure}[t]
\centering
{\footnotesize\textbf{(a)} Condensed SystemC RTL excerpt (\code{apb\_pmu\_timer\_subsys.h})\par}
\vspace{1.5pt}
\begin{lstlisting}[language=C++]
SC_MODULE(apb_gated_timer) { // Placed in PD_TIM via UPF
  sc_in<bool> clk, por_n, psel, penable, pwrite;
  sc_out<uint32_t> tim_rdata;  sc_out<bool> tim_irq;
  sc_signal<sc_uint<8>>  cfg_q;     // Retained (no por_n)
  sc_signal<sc_uint<16>> load_q;    // Retained shadow CSR
  sc_signal<sc_uint<16>> scratch_q, count_q; // Volatile
  void seq_proc() {
    if (!por_n.read()) { count_q.write(0); }
    else if (psel.read() && penable.read() && pwrite.read()) {
      if (paddr.read() == 0x10u)
        cfg_q.write(pwdata.read() & 3u);
      if (paddr.read() == 0x14u) load_q.write(pwdata.read());
    } else if (cfg_q.read() & 1u) {
      count_q.write(count_q.read()
                      ? count_q.read() - 1u : load_q.read());
    }
  }
  void comb_proc() { // mux_csr condenses 4-way paddr switch
    tim_rdata.write(mux_csr(paddr.read(),
                      cfg_q, load_q, scratch_q, count_q));
    tim_irq.write((cfg_q.read() & 3u) == 3u &&
                  count_q.read() == 1u);
  }
};
// Parent apb_pmu_timer_subsys::fabric_mux() (in PD_AON):
tim_psel.write(PSEL.read() && is_tim && !tim_iso_en.read());
const bool err = !is_pmu && (!is_tim || tim_iso_en.read());
if (PSEL.read() && PENABLE.read() && err) {
  PSLVERR.write(true);  PRDATA.write(0u);
} else {
  PSLVERR.write(false);
  PRDATA.write(is_tim ? tim_rdata.read() : pmu_rdata.read());
}
\end{lstlisting}
\vspace{2.5pt}
{\footnotesize\textbf{(b)} External power intent excerpt (\code{apb\_subsys.upf})\par}
\vspace{1.5pt}
\begin{lstlisting}
create_power_domain PD_AON -include_scope \
  -supply {primary SS_AON}
create_power_domain PD_TIM -elements {u_timer} \
  -supply {primary SS_TIM}
create_power_switch SW_TIM -domain PD_TIM \
  -input_supply_port {vin VDD_AON} \
  -output_supply_port {vout VDD_TIM} \
  -control_port {en u_pmu/tim_pwr_en_q} \
  -ack_port {ack tim_pwr_ack} \
  -on_state {ON vin {en}} -off_state {OFF {!en}}
set_isolation ISO_TIM -domain PD_TIM \
  -isolation_supply_set SS_AON -clamp_value 0 \
  -elements {u_timer/tim_irq u_timer/tim_rdata}
set_isolation_control ISO_TIM -domain PD_TIM \
  -isolation_signal u_pmu/tim_iso_en_q
set_retention RET_TIM -domain PD_TIM \
  -retention_supply_set SS_AON \
  -elements {u_timer/cfg_q u_timer/load_q}
set_retention_control RET_TIM -domain PD_TIM \
  -save_signal    {u_pmu/tim_ret_save_q posedge} \
  -restore_signal {u_pmu/tim_ret_restore_q posedge}
\end{lstlisting}
\vspace{-4pt}
\caption{Running example excerpted from \code{apb\_pmu\_timer\_subsys}. Zero low-power macros appear in the C++ RTL (a); all domain boundaries, switch controls, child-port clamps, and retention shadows reside in \code{apb\_subsys.upf} (b).}
\label{fig:motivating_code}
\end{figure}

Figure~\ref{fig:motivating_code} shows how those constructs interact on our reference design, \code{apb\_pmu\_timer\_subsys} (depicted structurally in Fig.~\ref{fig:arch}a). Always-on domain \code{PD\_AON} houses the software-programmable PMU (\code{u\_pmu}) and an APB3 fabric multiplexer (\code{fabric\_mux}); switchable domain \code{PD\_TIM} houses \code{u\_timer}, whose configuration and reload registers (\code{cfg\_q}, \code{load\_q}) are retained on \code{SS\_AON} while \code{scratch\_q} and down-counter \code{count\_q} lose state on power collapse. Walking through a single sleep--wake cycle on this excerpt reveals four failure modes that motivate Sections~\ref{sec:arch}--\ref{sec:formal}.

First, keeping \code{apb\_gated\_timer} free of power macros is required for multi-tool reuse, yet subtle to implement. Rewriting \code{cfg\_q} and \code{load\_q} with custom wrapper classes~\cite{gagarski2016scpower} makes the header unparseable by IEEE~1666.1 synthesis and formal front ends (\code{svcc}), whereas instrumenting it with TLM socket guards~\cite{mischkalla2013patmos} or energy monitors~\cite{vece2009pktool} leaves \code{seq\_proc} ticking and \code{count\_q} intact when \code{SW\_TIM} opens. Even the object names in Fig.~\ref{fig:motivating_code}b carry a trap: \code{ISO\_TIM} lists the child output ports \code{u\_timer/tim\_irq} and \code{u\_timer/tim\_rdata} (\code{sc\_out}, per IEEE~1801-2018 \S6.45), whereas \code{fabric\_mux} reads parent \code{sc\_signal} channels named \code{dut.tim\_irq} and \code{dut.tim\_rdata}. A simulator that binds UPF rules only to channel names misses the clamp without warning.

Second, suppose firmware cuts \code{tim\_pwr\_en\_q} before asserting \code{tim\_iso\_en\_q} (\code{LP\_BUG=1}). A four-state Verilog kernel corrupts \code{count\_q} to \code{16'hXXXX}, propagating \code{'X} onto \code{tim\_irq} and \code{IRQ}. Because \code{apb\_gated\_timer} uses synthesizable two-state \code{sc\_uint<16>} counters for arithmetic, zeroing powered-down state sets \code{count\_q = 0}; \code{count\_q == 1u} then evaluates to \code{false}, producing the exact \code{0} that \code{ISO\_TIM} (\code{-clamp\_value 0}) would have driven and masking the missing isolation. Inverting bits once on collapse (\code{\textasciitilde cur}) cancels out across an even number of rail glitches, and even a single random 16-bit draw per sleep window hits \code{count\_q == 1u} with probability $2^{-16}$. Section~\ref{sec:semantics} addresses sticky single-bit X-masking via per-clock \code{CORRUPT\_JITTER}, while Section~\ref{sec:formal} checks narrow $W$-bit comparators across all $2^W$ values via formal miters.

Third, power collapse breaks power-oblivious testbench components in both directions. If firmware pulses \code{tim\_ret\_save\_q} one cycle after \code{VDD\_TIM} drops (\code{LP\_BUG=2}) or reads volatile \code{scratch\_q} after cold wake before re-initializing it (\code{LP\_BUG=3}), a standard UVM register model still holds stale pre-sleep mirror expectations unless synchronized with retention and corruption events. Conversely, concurrent SVA checks monitoring \code{count\_q} inside \code{PD\_TIM} will misfire on scrambled values unless gated off while \code{VDD\_TIM == OFF}.

Fourth, directed simulation alone misses structural corner cases in Fig.~\ref{fig:motivating_code}a. Look at \code{fabric\_mux}: it forces \code{PRDATA = 0} on isolated timer accesses only when \code{PSEL \&\& PENABLE} holds. During the APB setup cycle (\code{PENABLE == 0}), the \code{else} branch forwards raw \code{tim\_rdata} onto \code{PRDATA}. Because APB monitors sample \code{PRDATA} only when \code{PENABLE \&\& PREADY}, directed tests never flag that setup-phase leak (\code{SUBSYS-02}). Likewise, C++ zero-initializes \code{sc\_uint<8> cfg\_q} at time zero; because \code{seq\_proc} omits \code{cfg\_q} from \code{!por\_n} to preserve timer configuration across warm sleep, \code{cfg\_q[1:0]} (\code{en}, \code{irq\_en}) powers up uninitialized at cold boot (\code{SUBSYS-03}). Formal checks expose both.

%% file: sections/architecture.tex
\section{Kernel Policy Hooks and UPF Runtime}
\label{sec:arch}

\subsection{Why External \code{force} Fails and How Policy Hooks Work}
\label{sec:hooks}

Executing \code{apb\_subsys.upf} on Fig.~\ref{fig:motivating_code}a without editing the RTL requires intercepting module scope, process dispatch, and signal updates. Before modifying the SystemC reference kernel, we evaluated an external prototype that registered stage callbacks and forced powered-down channels via \code{uvm\_hdl\_force}. That shortcut failed for three reasons. First, \code{SC\_METHOD} and \code{SC\_CTHREAD} processes inside a powered-down \code{SC\_MODULE} remained active on every clock edge, computing next-state values that fought the external force while wasting host CPU cycles. Second, driving a channel from an external observer tripped SystemC's \code{SC\_ONE\_WRITER} check unless single-writer enforcement was disabled globally across the design. Third, repurposing \code{force}/\code{release} overwrote legitimate fault-injection forces issued by UVM verification sequences.

We therefore introduced \code{sc\_power\_policy\_if.h}, a UPF-agnostic virtual policy interface (\code{+\SCHOOKDIFF{}/-16} kernel and channel diff lines plus the \code{\SCHOOKLOC{}}-line \code{sc\_power\_policy\_if.h} header in SystemC~3.0.3, with zero Tcl headers) spanning three kernel classes. On \code{sc\_module}, a nullable \code{sc\_domain\_policy\_if*} pointer associates each elaborated module subtree with its runtime domain without changing \code{SC\_MODULE} constructors. On \code{sc\_process\_b}, an orthogonal power-gate bit (\code{set\_power\_gated(bool)}) distinct from user \code{disable()} and \code{suspend()} suppresses static \code{SC\_METHOD} and \code{SC\_CTHREAD} dispatch while a domain is off; when power returns (\code{true $\rightarrow$ false}), the kernel re-schedules combinational methods and re-samples active reset levels so wakeup matches hardware power-on recovery. Finally, on \code{sc\_signal\_channel}, a nullable \code{sc\_signal\_power\_ext\_base* m\_power\_ext\_p} pointer (queried inline via \code{has\_active\_power\_policy()}, which stays \code{nullptr} whenever low-power simulation is inactive) intercepts \code{sc\_signal::update()}, \code{sc\_buffer}, \code{sc\_signal\_resolved}, and \code{sc\_signal\_rv<W>}. When an RTL driver process is gated off, \code{trigger\_power\_update()} queues the channel onto the kernel update list while preserving normal \code{SC\_ONE\_WRITER} enforcement, firing standard value-change and edge events across \code{bool}, C++ integers, \code{sc\_int}/\code{sc\_uint}, \code{sc\_bigint}/\code{sc\_biguint}, \code{sc\_bv}, \code{sc\_logic}, and \code{sc\_lv}. Because this interface knows nothing about UPF syntax, researchers prototyping custom power or fault-injection models can attach to \code{sc\_power\_policy\_if.h} without touching the kernel again.

\input{figures/fig_timeline}

\subsection{Avoiding Scheduler Warning \code{W553} Across Delta Cycles}
\label{sec:w553}

IEEE Std~1666-2023~\cite{ieee1666_2023} standardizes \code{sc\_stage\_callback\_if} hooks at \code{SC\_POST\_END\_OF\_ELABORATION}, \code{SC\_POST\_UPDATE}, \code{SC\_PRE\_TIMESTEP}, and \code{SC\_PRE\_STOP}. Our first runtime evaluated power switches and queued channel updates inside \code{SC\_POST\_UPDATE}. Running the SystemC~3.0.3 regression suite triggered kernel warning \code{(W553) forbidden action in stage callback}: IEEE~1666-2023 forbids immediate, delta, and timed \code{sc\_event::notify()} calls inside both \code{SC\_POST\_UPDATE} and \code{SC\_PRE\_TIMESTEP} so observers cannot mutate the runnable queue mid-transition. A second time-zero corner case surfaced on switch acknowledge ports (\code{-ack\_port}): when a switch enable started high prior to \code{sc\_start()}, seeding the internal \code{ack} driver cache to \code{true} suppressed the initial delta-0 write and left the RTL's \code{tim\_pwr\_ack} net stuck low.

Figure~\ref{fig:timeline} illustrates both a complete nine-clock sleep--wake waveform on \code{apb\_pmu\_timer\_subsys} and the intra-timestep schedule at collapse edge $C_3$ that avoids \code{W553}. At \code{SC\_POST\_END\_OF\_ELABORATION}, \code{systemc-lp} spawns an ungated kernel method (\code{\_\_sclp\_power\_eval\_\_}) sensitive to an internal \code{supply\_ev} event (triggered by \code{sc\_lp::upf::supply\_on/off}) and to every HDL control signal referenced by switches, isolation strategies, or retention rules. Switch \code{ack} caches start uninitialized so delta~0 always commits an explicit initial state. When an RTL write drops \code{tim\_pwr\_en\_q} in $\Delta_0$, its channel update wakes \code{\_\_sclp\_power\_eval\_\_} inside the legal \emph{evaluate} phase of $\Delta_1$. There, the engine solves the supply-switch network to a fixed point, evaluates isolation clamps \emph{before} collapsing the domain (ensuring \code{latch} clamps capture pre-collapse values), processes retention edges, updates \code{sc\_process\_b} gates, and calls \code{trigger\_power\_update()} so clamped and scrambled values commit in $\Delta_1$'s update phase. Once all delta cycles at that timestamp settle, \code{SC\_PRE\_TIMESTEP} executes strictly read-only PST legality checks, level-shifter threshold checks, and postponed FC4SC coverage sampling.

\begin{table}[t]
\caption{Low-Power Deactivation Modes in \code{systemc-lp}}
\label{tab:swap}
\centering
\footnotesize
\setlength{\tabcolsep}{4pt}
\begin{tabular}{@{}lcccc@{}}
\toprule
\textbf{Mode} & \textbf{How Selected} & \textbf{Recompile} & \textbf{Relink} & \textbf{\code{libtcl}} \\
\midrule
1a. CLI LP-ON & \code{./sim {-}{-}upf top.upf} & No & No & Linked \\
1b. CLI LP-OFF & \code{./sim} (or \code{{-}{-}no-lp}) & No & No & Linked \\
2. Link Stub & \code{SystemC::lp\_stub} & No & Yes & None \\
3. Elision & \code{-DSYSTEMC\_LP\_DISABLE} & Yes & Yes & None \\
\bottomrule
\end{tabular}
\end{table}

\subsection{Tcl Intent Evaluation, Port-Path Binding, and Deactivation}
\label{sec:swap}

Rather than writing a custom grammar for UPF, \code{systemc-lp} embeds \code{Tcl\_Interp} (\code{libtcl}) and registers handlers for 28 IEEE~1801 UPF~3.1 RTL commands plus two \code{systemc-lp} sign-off extensions (\code{add\_pst\_transition} and \code{set\_inrush\_limit}). Tcl variables, loops, conditionals, scoped \code{load\_upf -scope} calls, and wildcard \code{find\_objects} queries execute into a simulator-independent C++17 data model covering four functional groups: hierarchical scoping and supply networks (including \code{ONE\_HOT} and \code{PARALLEL} multi-driver \code{-resolve} nets); boolean power switches with delayed \code{-ack\_port} handshakes and concurrent parent-rail ramp limits (\code{set\_inrush\_limit}); named port states, composite power-state groups, PSTs, and directed \code{add\_pst\_transition} legality arcs; and boundary isolation, shadow retention with \code{-retention\_condition} hold checks, cross-voltage level shifters, and UPF~3.1 \code{sim\_corruption\_control} bring-up filters.

Binding that Tcl model onto an elaborated SystemC hierarchy must bridge the port-versus-channel split noted in Section~\ref{sec:overview}. When \code{ISO\_TIM} targets \code{u\_timer/tim\_irq}, that hierarchical path names an \code{sc\_out} port (\code{sc\_port\_base}), whereas clamping acts on the parent \code{sc\_signal\_channel} (\code{dut.tim\_irq}). During post-elaboration binding, \code{walk\_ports()} records every bound child port path on its underlying channel policy alongside the channel's own hierarchical name, forms the logical conjunction of all isolation strategies attached to a domain, and emits an \code{ELABORATION} diagnostic whenever any UPF element, switch port, or retention signal fails to resolve. Passing \code{{-}{-}lp-lint} runs static structural verification across all elaborated crossings and PST rows before cycle~0, catching missing isolation or level shifters, isolation or retention cells powered by their own collapsible domain rail, and unstaggered switch clusters sharing an inrush-limited supply without waiting for dynamic stimulus.

In daily regressions, teams rarely want separate build trees for functional and power-aware tests. Table~\ref{tab:swap} summarizes the three deactivation tiers supported by \code{systemc-lp}. Omitting \code{{-}{-}upf} on the command line (Mode~1b) unregisters per-timestep callbacks at elaboration and leaves \code{m\_power\_ext\_p == nullptr} on every channel in the same binary; Mode~2 relinks against the Tcl-free \code{SystemC::lp\_stub} archive on minimal build hosts; and Mode~3 compiles out all \code{sc\_lp} calls via \code{-DSYSTEMC\_LP\_DISABLE}. Section~\ref{sec:stack} measures the wall-clock cost of each mode against unhooked SystemC~3.0.3.

%% file: figures/fig_timeline.tex
\begin{figure*}[t]
\centering
\definecolor{aonblue}{RGB}{28,69,135}
\definecolor{timrust}{RGB}{175,65,25}
\definecolor{upfteal}{RGB}{0,115,115}
\definecolor{hazred}{RGB}{175,30,30}
\begin{tikzpicture}[
  x=1mm, y=1mm,
  font=\sffamily\scriptsize,
  >=Stealth,
  wf/.style={line width=0.65pt, draw=black!85},
  bus/.style={line width=0.55pt, draw=black!80, fill=white},
  jbus/.style={line width=0.55pt, draw=timrust, fill=timrust!8}
]

  \fill[timrust!9] (76,18) rectangle (124,63.5);
  \node[text=timrust, font=\sffamily\tiny\bfseries] at (100,61.8) {PD\_TIM = CORRUPT ($K=3$ off-domain clocks)};

  \foreach \i/\lbl in {0/C_0,1/C_1,2/C_2,3/C_3,4/C_4,5/C_5,6/C_6,7/C_7,8/C_8} {
    \pgfmathsetmacro{\xx}{28 + \i*16}
    \draw[ densely dotted, draw=black!30, line width=0.4pt] (\xx,18) -- (\xx,60);
    \node[font=\sffamily\tiny, text=black!70] at (\xx+8,65) {$\lbl$};
  }

  \node[anchor=east, font=\sffamily\scriptsize\bfseries] at (26,56.7) {\code{PCLK}};
  \draw[wf] (28,55)
    \foreach \i in {0,...,7} {
      -- ({28+\i*16},58.2) -- ({36+\i*16},58.2) -- ({36+\i*16},55) -- ({44+\i*16},55)
    };

  \node[anchor=east, font=\sffamily\scriptsize] at (26,50.2) {\code{save / restore}};
  \draw[wf, draw=upfteal] (28,48.5) -- (44,48.5) -- (44,51.8) -- (60,51.8) -- (60,48.5)
    -- (140,48.5) -- (140,51.8) -- (156,51.8) -- (156,48.5) -- (164,48.5);
  \node[font=\sffamily\tiny, text=upfteal] at (52,53) {\code{save}$\uparrow$};
  \node[font=\sffamily\tiny, text=upfteal] at (148,53) {\code{restore}$\uparrow$};

  \node[anchor=east, font=\sffamily\scriptsize] at (26,43.7) {\code{tim\_iso\_en}};
  \draw[wf, draw=aonblue] (28,42) -- (60,42) -- (60,45.3) -- (156,45.3) -- (156,42) -- (164,42);

  \node[anchor=east, font=\sffamily\scriptsize, text=timrust] at (26,37) {\code{VDD\_TIM}};
  \draw[wf, draw=timrust] (28,38.8) -- (76,38.8) -- (78.5,35.2) -- (124,35.2) -- (126.5,38.8) -- (164,38.8);
  \node[font=\sffamily\tiny, text=timrust] at (42,37) {0.90V};
  \node[font=\sffamily\tiny, text=timrust] at (100,37) {0.0V (OFF)};
  \node[font=\sffamily\tiny, text=timrust] at (145,37) {0.90V};

  \node[anchor=east, font=\sffamily\scriptsize] at (26,30.5) {\code{load\_q} {\tiny(ret.)}};
  \draw[bus] (28,28.8) rectangle (76,32.2);
  \node[font=\sffamily\tiny] at (52,30.5) {\code{0x0024}};
  \draw[bus, fill=black!7, densely dashed] (76,28.8) rectangle (140,32.2);
  \node[font=\sffamily\tiny, text=upfteal] at (108,30.5) {shadow holds \code{0x0024}};
  \draw[bus, draw=upfteal, fill=upfteal!8] (140,28.8) rectangle (164,32.2);
  \node[font=\sffamily\tiny, text=upfteal] at (152,30.5) {\code{0x0024}};

  \node[anchor=east, font=\sffamily\scriptsize] at (26,24) {\code{count\_q} {\tiny(jitter)}};
  \draw[bus] (28,22.3) rectangle (76,25.7);
  \node[font=\sffamily\tiny] at (52,24) {\code{0x0003..0x0001}};
  \draw[jbus] (76,22.3) rectangle (92,25.7);
  \node[font=\sffamily\tiny, text=timrust] at (84,24) {\code{0x9E37}};
  \draw[jbus] (92,22.3) rectangle (108,25.7);
  \node[font=\sffamily\tiny, text=timrust] at (100,24) {\code{0x4A1C}};
  \draw[jbus] (108,22.3) rectangle (124,25.7);
  \node[font=\sffamily\tiny, text=timrust] at (116,24) {\code{0xD08B}};
  \draw[bus] (124,22.3) rectangle (164,25.7);
  \node[font=\sffamily\tiny] at (144,24) {\code{0x0000} (\code{por\_n=0})};

  \node[anchor=east, font=\sffamily\scriptsize] at (26,19.3) {\code{dut.IRQ}};
  \draw[wf, draw=aonblue] (28,18.4) -- (164,18.4);
  \draw[densely dashed, draw=hazred, line width=0.6pt] (76,18.4) -- (76,21.0) -- (92,21.0) -- (92,18.4);
  \node[anchor=west, font=\sffamily\tiny, text=hazred] at (92.5,19.8) {leak if \code{ISO\_TIM} missing (\code{LP\_BUG=1})};

  \draw[->, draw=timrust, line width=0.55pt] (76,18) -- (34,13.2);
  \draw[draw=black!65, fill=black!3, rounded corners=1pt] (2,1.5) rectangle (174,13.2);
  \node[anchor=west, font=\sffamily\tiny\bfseries] at (3.5,10.8) {Intra-Timestep Schedule at $C_3$ (Avoids IEEE~1666-2023 \code{W553} Stage-Callback Violation):};

  \node[draw=black!75, fill=white, rounded corners=0.8pt, minimum width=27mm, minimum height=5.8mm, font=\sffamily\tiny, align=center] (s1) at (18,5.2) {\textbf{Eval ($\Delta_0$)}\\[0.3pt]RTL \code{pwr\_en <= 0}};
  \node[draw=black!75, fill=white, rounded corners=0.8pt, minimum width=24mm, minimum height=5.8mm, font=\sffamily\tiny, align=center] (s2) at (49,5.2) {\textbf{Update ($\Delta_0$)}\\[0.3pt]\code{pwr\_en} commits};
  \node[draw=aonblue, fill=aonblue!6, rounded corners=0.8pt, minimum width=43mm, minimum height=5.8mm, font=\sffamily\tiny, align=center] (s3) at (88,5.2) {\textbf{Eval ($\Delta_1$): \code{\_\_sclp\_power\_eval\_\_}}\\[0.3pt]Solves \code{SW\_TIM}, gates \code{u\_timer}, queues update};
  \node[draw=timrust, fill=timrust!7, rounded corners=0.8pt, minimum width=31mm, minimum height=5.8mm, font=\sffamily\tiny, align=center] (s4) at (130.5,5.2) {\textbf{Update ($\Delta_1$)}\\[0.3pt]Commits clamp \& jitter};
  \node[draw=upfteal, fill=upfteal!7, rounded corners=0.8pt, minimum width=22mm, minimum height=5.8mm, font=\sffamily\tiny, align=center] (s5) at (161,5.2) {\textbf{\code{PRE\_TIMESTEP}}\\[0.3pt]Read-only PST/CG};

  \draw[->, line width=0.5pt] (s1.east) -- (s2.west);
  \draw[->, line width=0.5pt] (s2.east) -- (s3.west);
  \draw[->, line width=0.5pt] (s3.east) -- (s4.west);
  \draw[->, line width=0.5pt] (s4.east) -- (s5.west);

\end{tikzpicture}
\caption{Sleep--wake execution on \code{apb\_pmu\_timer\_subsys} ($C_0\dots C_8$). While retained \code{load\_q} holds \code{0x0024} on \code{SS\_AON}, volatile \code{count\_q} re-scrambles on every active clock edge inside \code{CORRUPT} (\code{CORRUPT\_JITTER}). Bottom: running \code{\_\_sclp\_power\_eval\_\_} in the evaluate phase ($\Delta_1$) rather than \code{SC\_POST\_UPDATE} keeps channel updates legal without triggering kernel warning \code{W553}.}
\label{fig:timeline}
\end{figure*}
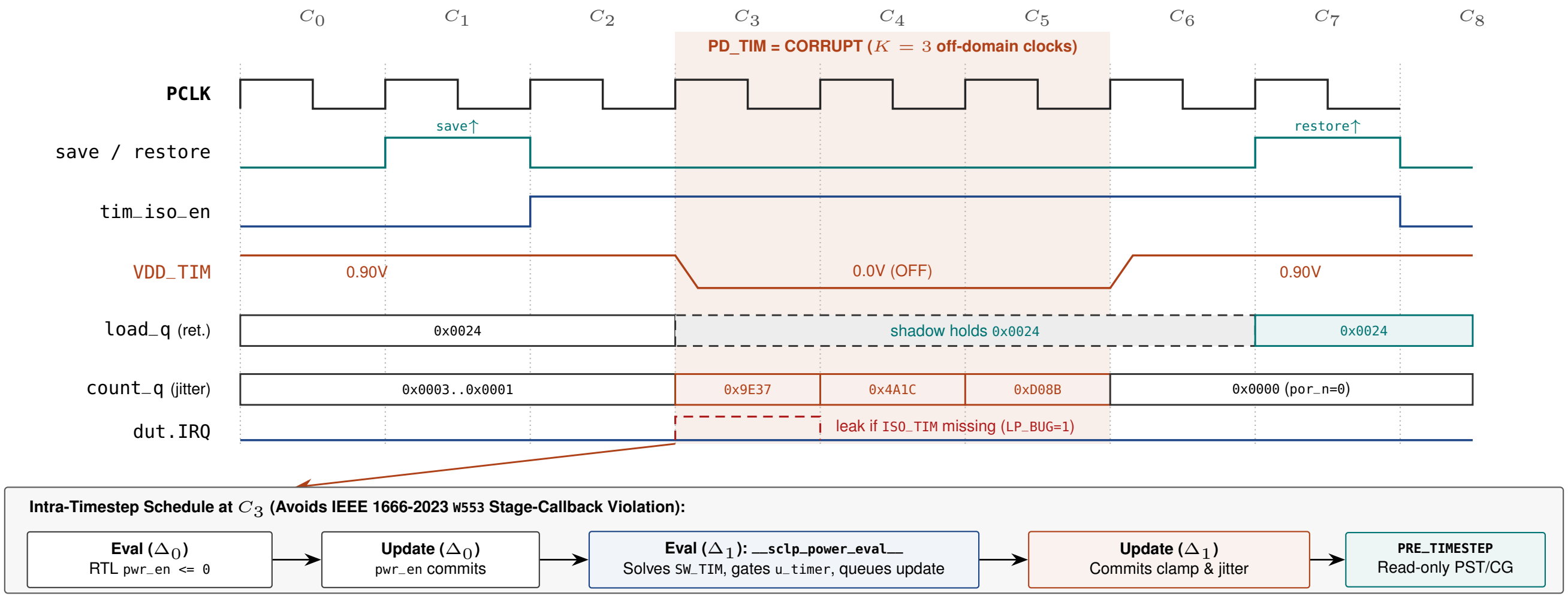

%% file: sections/semantics_and_dv.tex
\section{Power-Collapse Semantics, Verilator Binding, and Testbench Adapters}
\label{sec:semantics}

\subsection{Four-State \code{'X}/\code{'Z} and Two-State Jitter Corruption}
\label{sec:jitter}

SystemC supports both four-state logic vectors and two-state hardware integers, and \code{systemc-lp} handles both according to each channel's type. On four-state signals (\code{sc\_logic}, \code{sc\_lv<W>}), \code{sc\_power\_policy\_if.h} drives \code{SC\_LOGIC\_X} across powered-down channels and \code{SC\_LOGIC\_Z} under \code{-clamp\_value Z} on resolved multi-driver buses (\code{sc\_signal\_resolved}, \code{sc\_signal\_rv<W>}), where \code{evaluate\_driver\_power\_action()} overrides only the isolated or collapsed driver's contribution prior to resolution so active peers keep driving valid levels without false contention.

In synthesizable RTL, however, \code{sc\_lv<W>} defines bitwise logic without arithmetic operators (\code{+}, \code{-}, \code{<}), so IEEE~1666.1 high-level synthesis models and \code{svcc} write datapaths and counters with two-state types (\code{bool}, \code{uint32\_t}, \code{sc\_int<W>}, \code{sc\_uint<W>}), while \code{Verilator {-}{-}sc} compiles SystemVerilog \code{logic} down to two-state C++ integers for simulation speed. Static two-state corruption schemes fail in predictable ways. Driving \code{ALL\_ZEROS} (\code{0}) matches active-high idle levels and AND-clamp values, masking missing isolation on interrupt or valid lines as seen on \code{tim\_irq} in Section~\ref{sec:overview}. Inverting the pre-collapse word (\code{FLIP\_CURRENT}, \code{\textasciitilde cur}) exposes idle zeros on a single shutdown, yet still fails if a control bit happened to be \code{1} right before power-off (flipping \code{1 $\rightarrow$ 0} mimics an AND-clamp) or if an even number of rapid rail glitches exploits bitwise involution ($\neg\neg v = v$) to restore the original state. Drawing one pseudo-random word per collapse epoch via \code{SEEDED\_SCRAMBLE} (\code{splitmix64($\text{seed} \oplus \text{hash}(\text{path}) \oplus \text{epoch}$)}) avoids deterministic fixed points, but holds that word static throughout sleep: an unisolated 1-bit control still lands on its benign value in $50\%$ of single-sleep runs.

\input{figures/fig_escape}

One draw per sleep window is not enough when clocks keep ticking in active sink domains. Under \code{CORRUPT\_JITTER} (\code{{-}{-}lp-corrupt=jitter}), while a domain remains in \code{CORRUPT} across $K$ active clock edges, \code{systemc-lp} increments \code{jitter\_step} and re-hashes \code{splitmix64($\text{seed} \oplus \text{hash}(\text{path}) \oplus \text{epoch} \oplus \text{jitter\_step}$)} across unisolated boundary channels and volatile registers (visible on \code{count\_q} across $C_3\dots C_5$ in Fig.~\ref{fig:timeline}). For an unisolated 1-bit control \emph{continuously observed} by a sticky always-on sink (such as an interrupt latch, asynchronous set/reset pin, or FIFO push enable), every off-domain edge becomes an independent trial, driving the masking escape probability down exponentially:
\begin{equation}
P_{\text{escape}}(K) = 2^{-K}.
\label{eq:jitter}
\end{equation}
Two structural cases remain outside Eq.~\eqref{eq:jitter} and motivate combining simulation with formal verification (Fig.~\ref{fig:escape}). First, if an active sink strobes an unisolated 1-bit net on only $M \ll K$ cycles (for instance, during sporadic bus reads), escape scales as $2^{-M}$. Second, when an unisolated $W$-bit bus feeds an equality comparator such as \code{count\_q == 1u} ($W = 16$) or a 32-bit tag decoder, each jitter step hits the trigger word with probability $2^{-W}$, so $(1 - 2^{-W})^K \approx 1$ across practical simulation windows. Section~\ref{sec:formal} discharges those narrow-cone paths symbolically with $P_{\text{escape}} = 0$.

\subsection{Stateful Simstates, Retention Brownout, and Dual-Rail SRAMs}
\label{sec:sram}

To execute all four IEEE~1801 \code{simstate} behaviors accurately, \code{systemc-lp} inspects process sensitivity lists at elaboration, classifying edge-sensitive processes and \code{SC\_CTHREAD}s as sequential and data-sensitive \code{SC\_METHOD}s as combinational. Entering \code{CORRUPT\_ON\_ACTIVITY} scrambles sequential registers while holding combinational outputs until a domain input toggles; entering drowsy \code{CORRUPT\_STATE\_ON\_CHANGE} (via \code{supply\_partial\_on} or \code{add\_power\_state}) leaves processes runnable so any write or ungated clock toggle raises \code{RET\_HAZARD} and scrambles the disturbed register. During incremental SoC bring-up, UPF~3.1 \code{sim\_corruption\_control} (\code{ALL}, \code{REGISTER\_ONLY}, \code{COMBINATIONAL\_ONLY}, \code{NONE}) and \code{set\_simstate\_behavior} let engineers scope collapse semantics down to individual domains or black-box macros.

Retention and SRAM sign-off use continuous supply and control checks. Reusable UPF~3.1 \code{set\_retention\_elements} lists expand across SystemC channels, Verilator flip-flops, UVM register models, and memory blocks. Throughout sleep, the strategy evaluator checks both boolean \code{-retention\_condition} hold expressions (such as gated retention clocks) and the minimum standby voltage declared on \code{add\_port\_state}: if a standby rail remains \code{PARTIAL\_ON} yet droops below its retention voltage floor, or if an active reset clobbers a register right after \code{restore}, \code{systemc-lp} invalidates the shadow value, scrambles the target, and reports \code{RET\_HAZARD} or \code{RET\_CLOBBERED\_BY\_RESET}. For embedded SRAMs with separate periphery (\code{VDD\_PERIPH}) and bitcell-array (\code{VDDM\_ARRAY}) rails, \code{sclp\_memory.h} attaches dual-rail policies to C++ arrays, \code{std::array}, and Verilator \code{VlUnpacked} buffers: collapsing \code{VDDM\_ARRAY} scrambles the array at host word speed, whereas collapsing \code{VDD\_PERIPH} while \code{VDDM\_ARRAY} stays on snapshots the array and checks whether unisolated write-enable glitches disturb any word during periphery shutdown.

\subsection{Automatic \code{Verilator {-}{-}sc} Binding and \code{package UPF} DPI}
\label{sec:verilator}

Under \code{Verilator {-}{-}sc}, module boundary ports elaborate as \code{sc\_in}/\code{sc\_out} channels, but internal SystemVerilog registers and unpacked arrays flatten into plain C++ struct fields inside \code{V<prefix>\_\_\_024root} (for example, \code{dut\_\_DOT\_\_u\_tim\_\_DOT\_\_cfg\_q}). Requiring manual C++ registration calls per flip-flop would break unmodified-RTL portability. Instead, our CMake integration (\code{SystemCLPVerilator.cmake}) runs \code{gen\_verilator\_bind.py} after Verilation: the script scans \code{V<prefix>\_\_\_024root.h} to decode \code{\_\_DOT\_\_} and \code{\_\_BRA\_\_..KET\_\_} symbols back into IEEE~1800 hierarchical paths, cross-references delayed nonblocking temporaries (\code{\_\_VdlyVal\_\_*}) and sequential update functions in \code{V<prefix>\_\_\_024root*.cpp} to separate flip-flops from combinational nets, and emits a single \code{sclp\_bind\_verilated(dut)} header so UPF \code{find\_objects} globs resolve across SystemVerilog RTL. In mixed-language testbenches, companion package \code{systemc\_lp\_upf\_pkg.sv} implements IEEE~1801's \code{package UPF} (\code{supply\_on}, \code{supply\_off}, \code{supply\_partial\_on}, \code{query\_supply}, and \code{get\_domain\_simstate}) over DPI-C exports backed by the same supply solver.

\subsection{Synchronizing Power-Oblivious Testbenches}
\label{sec:bridges}

Core \code{libsystemc-lp} links against none of \code{uvm-systemc}, \code{fc4sc}, or \code{crave}, and none of those upstream Accellera libraries were modified for low-power support. Instead, optional header-only adapters (\code{<sysc\_lp/bridges/*.h>}, exported as CMake \code{INTERFACE} targets) repair the testbench assumptions that break when hardware domains power down.

At the register and bus layers, \code{bridges/uvm.h} walks \code{uvm\_reg\_block::get\_root\_blocks()} after elaboration, maps each \code{uvm\_reg} HDL path onto its resolved UPF domain and retention rule, and attaches register callbacks: reading a volatile register after cold wake prior to software re-initialization raises \code{RAL\_CORRUPT}, whereas retained registers snapshot \code{get\_mirrored\_value()} on \code{RETENTION\_SAVE}, restore \code{predict(saved\_mirror)} on \code{RETENTION\_RESTORE}, and invalidate mirrors on \code{RETENTION\_HAZARD}. On loosely-timed TLM-2.0 interconnects, initiators routinely cache raw host memory pointers returned by \code{get\_direct\_mem\_ptr()}, bypassing \code{b\_transport} on later accesses; wrapping a target socket in \code{sclp\_tlm\_domain\_guard} (\code{bridges/tlm.h}) invokes \code{invalidate\_direct\_mem\_ptr()} upon domain collapse and rejects transport accesses while off. For assertions, coverage, and stimulus, \code{bridges/fc4sc\_sva.h} wraps functional checks in \code{SCLP\_ASSERT\_DOMAIN\_PROPERTY} so \code{CORRUPT\_JITTER} scrambling disables off-domain SVA threads instead of causing false failures, synthesizes Preponed isolation assertions and legal PST state/transition covergroups (honoring \code{add\_pst\_transition} exclusions) that export \code{UCIS.xsd}-valid XML at \code{SC\_PRE\_STOP}, and pairs with \code{bridges/crave.h} so constrained-random PMU sequences sweep legal PST arcs.

%% file: figures/fig_escape.tex
\begin{figure}[t]
\centering
\definecolor{aonblue}{RGB}{28,69,135}
\definecolor{timrust}{RGB}{175,65,25}
\definecolor{upfteal}{RGB}{0,115,115}
\definecolor{hazred}{RGB}{175,30,30}
\begin{tikzpicture}
\begin{semilogyaxis}[
  width=0.98\columnwidth,
  height=4.7cm,
  xlabel={Off-domain clock edges $K$ while \code{simstate == CORRUPT}},
  ylabel={Masking escape prob.\ $P_{\text{escape}}(K)$},
  xmin=1, xmax=16,
  ymin=1e-5, ymax=2.2,
  xtick={1,4,8,12,16},
  ytick={1,1e-1,1e-2,1e-3,1e-4,1e-5},
  grid=both,
  grid style={dotted, black!25},
  label style={font=\sffamily\scriptsize},
  tick label style={font=\sffamily\tiny},
  legend style={
    at={(0.02,0.03)},
    anchor=south west,
    font=\sffamily\tiny,
    fill=white,
    fill opacity=0.92,
    draw=black!40,
    inner sep=1.5pt,
    cells={anchor=west}
  }
]
  \addplot[draw=black!55, dashed, line width=0.75pt, mark=o, mark size=1.4pt]
    coordinates {(1,1) (4,1) (8,1) (12,1) (16,1)};
  \addlegendentry{Static \code{0} / even-glitch \code{\textasciitilde cur} ($1.0$)}

  \addplot[draw=hazred, dashdotted, line width=0.85pt, mark=triangle*, mark size=1.5pt]
    coordinates {(1,0.9999999998) (4,0.9999999991) (8,0.9999999981) (12,0.9999999972) (16,0.9999999963)};
  \addlegendentry{\code{CORRUPT\_JITTER}, 32-bit \code{==} ($\approx 1$)}

  \addplot[draw=timrust, densely dashed, line width=0.8pt, mark=square*, mark size=1.3pt]
    coordinates {(1,0.5) (4,0.5) (8,0.5) (12,0.5) (16,0.5)};
  \addlegendentry{\code{SEEDED\_SCRAMBLE}, 1-bit ($0.5$)}

  \addplot[draw=aonblue, solid, line width=0.95pt, mark=*, mark size=1.4pt, domain=1:16, samples=16]
    {2^(-x)};
  \addlegendentry{\code{CORRUPT\_JITTER}, sticky 1-bit ($2^{-K}$)}

  \node[anchor=north east, font=\sffamily\tiny\bfseries, text=upfteal, fill=upfteal!7, draw=upfteal, rounded corners=1pt, inner sep=1.5pt]
    at (axis cs:15.7,0.22) {2-copy formal miter (\code{svcc}/\code{InFormal}): $P_{\text{escape}}=0$};
\end{semilogyaxis}
\end{tikzpicture}
\caption{Functional X-masking escape probability $P_{\text{escape}}(K)$ for an unlinted, unisolated two-state net across $K$ off-domain clock edges (\code{{-}{-}lp-lint}, UPF monitors, and four-state \code{'X}/\code{'Z} are deterministic). \code{CORRUPT\_JITTER} drives sticky 1-bit escape down as $2^{-K}$, whereas wide equality decoders (\code{tag == 32'hDEAD\_BEEF}) stay near $1.0$ and require the symbolic two-copy miters in Section~\ref{sec:formal}.}
\label{fig:escape}
\end{figure}
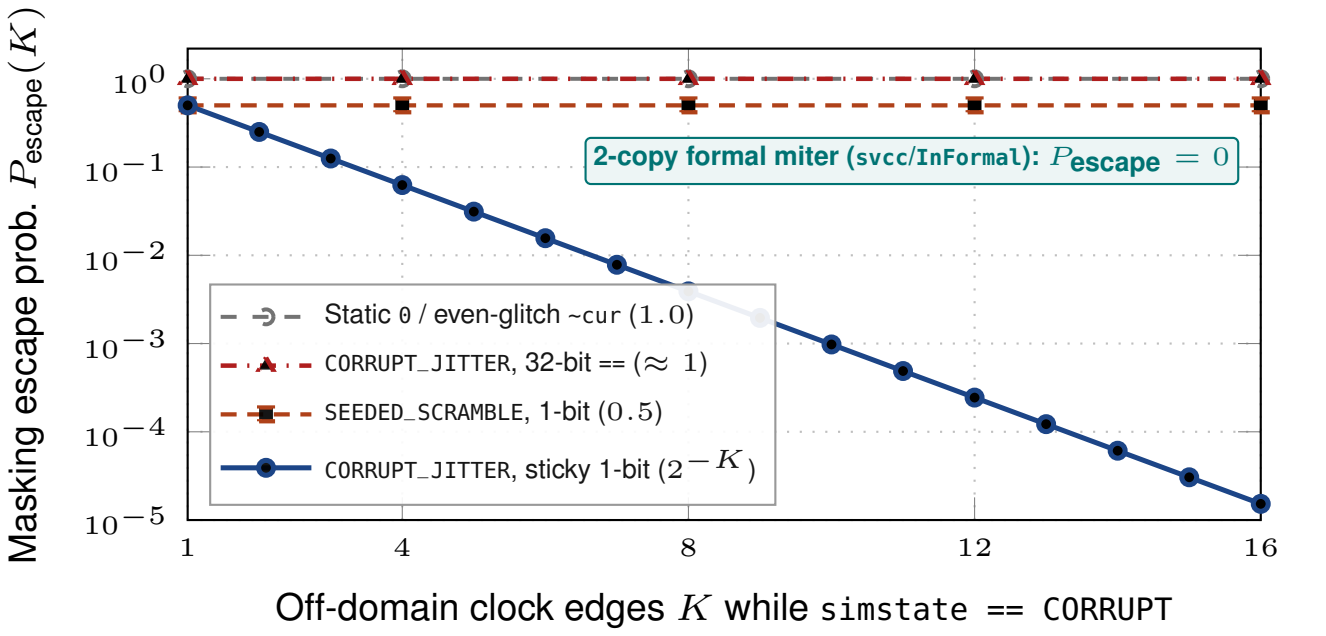

%% file: sections/formal_continuity.tex
\section{Shared-Source Word-Level Formal Verification}
\label{sec:formal}

\begin{table*}[t]
\caption{Complementary Scope and Soundness Boundaries of Dynamic \code{systemc-lp} Simulation vs.\ Formal \code{svcc} + \code{InFormal} Verification}
\label{tab:formal_vs_sim}
\centering
\footnotesize
\setlength{\tabcolsep}{5pt}
\begin{tabularx}{\textwidth}{@{}>{\raggedright\arraybackslash}p{3.4cm}>{\raggedright\arraybackslash}X>{\raggedright\arraybackslash}X@{}}
\toprule
\textbf{Verification Dimension} & \textbf{Dynamic \code{systemc-lp} (SystemC / \code{Verilator {-}{-}sc})} & \textbf{Formal \code{svcc prove-upf} + \code{InFormal} (Paths A \& B)} \\
\midrule
Off-domain corruption coverage & \textbf{Four-state \code{'X} \& 2-state jitter:} $P_{\text{escape}} = 2^{-K}$ on sticky 1-bit controls under \code{CORRUPT\_JITTER}, but $(1 - 2^{-W})^K \approx 1$ on $W$-bit comparators. & \textbf{Exhaustive ($2^W$ states):} \code{iso-2copy} and \code{check\_lp} expose 32-bit equality decoders and APB setup-phase mux leakage in one SAT/SMT query. \\
\addlinespace[2pt]
Post-wakeup X-leakage \& boot & \textbf{Stimulus-dependent:} requires directed read or IRQ checks before re-write; C++ zero-init masks cold-boot resetless flip-flops. & \textbf{Sequential 2-copy \& structural lint:} \code{formal\_lint} flags resetless control flops (\code{SV-FPV-002}); \code{check\_lp} isolates post-wakeup corruption leaks. \\
\addlinespace[2pt]
Unisolated cones \& PMU rules & \textbf{Concrete traces:} exercises programmed firmware sequences; unvisited CSR write combinations remain untested. & \textbf{Dead-cone waivers \& 2QBF contracts:} prunes unreachable cones (\code{svcc\_\_ub}) via COI analysis and validates firmware assumptions via \code{proposal\_validate}. \\
\addlinespace[2pt]
Clocking \& asynchronous resets & \textbf{Event-driven timing:} executes multi-clock CDC wakeup handshakes, sub-cycle async reset/restore races, and SystemC delta cycles. & \textbf{Single-clock sampled-reset model:} Btor2 lowering assumes one root clock, samples async resets at clock edges, and abstracts intra-step delta ordering. \\
\addlinespace[2pt]
Dual-rail SRAMs ($N \times W$ bits) & \textbf{$O(N)$ host word speed:} scrambles and write-disturb-checks full memory arrays during firmware and DMA sequences. & \textbf{State-space scaling:} large bitcell arrays overwhelm bit-level IC3 unless memories are blackboxed or abstracted to symbolic locations. \\
\bottomrule
\end{tabularx}
\end{table*}

\subsection{Word-Level UPF Lowering and Miters in \code{svcc}}
\label{sec:lowering}

As Fig.~\ref{fig:escape} showed, multi-cycle jitter drives sticky single-bit X-masking down as $2^{-K}$ in simulation, yet wide equality decoders, setup-phase bus multiplexers ignored by protocol monitors, and cold-boot reset omissions demand symbolic proof (Table~\ref{tab:formal_vs_sim}). Because \code{systemc-lp} keeps \code{SC\_MODULE} headers strictly within the IEEE~1666.1 synthesizable subset~\cite{ieee1666_1_2016} with zero power macros, we extended \code{svcc}~\cite{kubavat2026svcc} to read the same C++17 header and external \code{.upf} specification.

\code{svcc} first lifts \code{SC\_MODULE} hierarchies under exact LP64 C++17 integer promotion, width truncation, and undefined-behavior (\code{svcc\_\_ub}) semantics, then evaluates a 22-command structural subset of IEEE~1801 UPF~3.1 in pure Rust (\code{svcc-ir::upf}, with \code{load\_upf -scope} and \code{find\_objects} globs); it expands Tcl \code{set} variables and \code{foreach} loops and rejects procedural Tcl constructs (\code{if}, \code{proc}, \code{expr}) fail-closed. Static pass \code{upf-static} verifies that every \code{-elements} path, switch control, and cross-voltage shifter resolves against the lifted module ports before lowering begins.

During structural lowering (\code{svcc-passes::upf\_lower}), each gated child instance is specialized with explicit power-intent hardware. Every isolated child output port $y$ of domain $D$ receives a boundary multiplexer driven by strategy activation $\text{act}_i$, clamp constant or latch $c_{\text{val}}$, domain supply predicate $\text{on}(D)$, and a fresh symbolic primary input $\code{upf\_\_float\_}y$ modeling off-domain floating corruption:
\begin{equation}
y_{\text{iso}} = \text{act}_i \mathrel{?} c_{\text{val}} : \bigl(\text{on}(D) \mathrel{?} y_{\text{raw}} : \code{upf\_\_float\_}y\bigr).
\label{eq:isomux}
\end{equation}
Domain-wide isolation is formed conjunctively as $\code{upf\_\_off\_isolated\_}D \triangleq \neg\text{on}(D) \land \bigwedge_i \text{act}_i$. Each retained register $r \in D$ receives an always-on shadow register updated on \code{save} when $\text{on}(D) \land \text{ret}(D)$ and copied back on \code{restore}, whereas non-retained registers load unconstrained \code{upf\_\_corrupt\_*} inputs whenever $\neg\text{on}(D)$. Because \code{upf\_lower} keeps multi-bit ports, isolation multiplexers, and shadow registers as word-level $W$-bit vector operations rather than bit-blasted gate cells, formal elaboration stays compact; Guimar\~{a}es et al.~\cite{guimaraes2026vectorization} recently showed across 1,157 Verilog benchmarks that preserving vector structure cuts elaboration time by 28\% and memory footprint by 51\% ahead of verification and synthesis.

From that word-level representation, \code{svcc prove-upf} compiles word-level Btor2~\cite{niemetz2018btor2} models into SMT-LIB2 \code{QF\_BV} proof obligations solved by Z3~\cite{demoura2008z3}: \code{cpp-ub} proves C++17 undefined-behavior freedom (\code{svcc\_\_ub == 0}); \code{iso-2copy} builds a two-copy self-composition miter ($M^{(0)} \times_{\mathcal{I}} M^{(1)}$~\cite{yang2018lazyself,tan2026secic3}, Fig.~\ref{fig:arch}b) driven by shared active inputs $X$ and independent off-domain corruption vectors $\xi^{(0)}, \xi^{(1)}$ under \code{upf\_\_off\_isolated\_$D$ == 1} to prove output non-interference across all $2^W$ states; \code{retention} proves $\code{Save} \rightarrow \code{Collapse}(\xi) \rightarrow \code{Wake+Restore}$ recovery from unconstrained initial states; and \code{proto-bmc} checks protocol violation flags such as \code{ISO\_WHILE\_OFF} on self-supplied isolation cells.

\subsection{Dual-Path Model Checking and Firmware Contracts in \code{InFormal}}
\label{sec:threeproofs}

To cross-check \code{svcc}'s Btor2 miters against an independent model checker and reason over multi-step PMU sequences, we pair \code{svcc} with \code{InFormal}~\cite{kubavat2026informal} along two paths. In \textbf{Path~A (direct UPF model checking)}, \code{svcc sc2sv} translates the unmodified C++ RTL into synthesizable SystemVerilog, and \code{InFormal} reads \code{design.upf} alongside it: \code{formal\_lint} audits structural reset coverage (\code{SV-FPV-002}), \code{load\_upf} encodes the PST into CNF checked by \code{InFormal}'s CaDiCaL-backed~\cite{biere2024cadical} SAT solver, \code{check\_lp} proves active outputs immune to off-domain corruption while pruning structurally dead unisolated cones (\code{formally\_waived}) via cone-of-influence analysis, and \code{lp\_diagnose} extracts minimal backward causal signal slices whenever post-wakeup volatility leaks to an output.

In \textbf{Path~B (lowered-SVA induction and firmware contract validation)}, \code{svcc sc2sv {-}{-}upf} emits SystemVerilog containing explicit Eq.~\eqref{eq:isomux} multiplexers, shadow registers, witness covers, and concurrent SVA assertions. \code{InFormal} proves those safety invariants by $k$-induction (alongside CaDiCaL-backed bounded model checking and optional binary AIGER witness emission for Certifaiger~\cite{yu2021certifaiger}). Because hardware PMUs such as \code{u\_pmu} expose raw power, isolation, and retention CSR bits to software, unconstrained formal primary inputs trigger illegal single-write shutdowns. Path~B resolves this through \code{InFormal}'s \code{proposal\_validate} gate, which checks candidate primary-input SVA assumptions for bounded consistency, 2QBF receptiveness (confirming the environment cannot be forced into a deadlock), witness-cover preservation, and non-vacuity. Once validated, that CSR write-ordering contract completes the inductive hardware proof and can be reused as a dynamic FC4SC monitor on firmware sequences in \code{systemc-lp}, while \code{systemc-lp} exercises the full-size dual-rail SRAM arrays and multi-clock wakeup handshakes that lie beyond single-clock bit-level IC3 (Table~\ref{tab:formal_vs_sim}).

%% file: sections/evaluation.tex
\section{Empirical Evaluation}
\label{sec:eval}

We evaluate four aspects of the flow across our open C++17 and Rust verification stack: kernel patch footprint and deactivation overhead against unhooked SystemC~3.0.3 (Section~\ref{sec:stack}); dynamic fault-injection coverage across SystemC, \code{Verilator {-}{-}sc}, SRAMs, and testbench adapters (Section~\ref{sec:case_sc}); word-level formal proof speed and narrow-comparator mutation sensitivity (Section~\ref{sec:formal_eval}); and defects uncovered in both \code{systemc-lp} and the APB subsystem by cross-checking identical files across tools (Section~\ref{sec:bug_discovery}), followed by scope boundaries (Section~\ref{sec:limitations}).

\subsection{Experimental Setup, Footprints, and Turnoff Cost}
\label{sec:stack}

Timing measurements were collected on an Apple M5~Max workstation (36\,GB unified RAM, AppleClang~21.0.0, \code{rustc}~1.100.0, Verilator~5.052, Z3~5.3.3, and CaDiCaL~3.0.1), with C++ build and non-Verilator regression gates ($29/29$ \code{systemc-lp} CTests) also verified on \code{x86\_64} Linux (\code{GCC}~15.2.0, CMake~4.2). Table~\ref{tab:stack} summarizes repository provenance, footprints, and regression gates. All five C++ repositories pass their Release regression suites (\code{systemc-lp} compiled with \code{-Wall -Wextra -Werror}), and both the hooked \code{systemc} kernel and \code{systemc-lp} pass \code{ASan+UBSan} (\code{UBSAN\_OPTIONS=halt\_on\_error=1}) suites with zero memory leaks, while all generated UCIS~1.0 coverage reports validate against Accellera's \code{UCIS.xsd} schema.

\begin{table}[t]
\caption{Component Inventory, Footprints, and Verified Regression Gates}
\label{tab:stack}
\centering
\scriptsize
\setlength{\tabcolsep}{2.8pt}
\begin{tabularx}{\columnwidth}{@{}l>{\raggedright\arraybackslash}Xrr@{}}
\toprule
\textbf{Repo (SHA)} & \textbf{Standard / Role} & \textbf{Footprint} & \textbf{Verified Gate} \\
\midrule
\code{systemc} (\code{\SCCOMMIT}) & 1666-2023 + LP hooks & +\SCHOOKDIFF{}+\SCHOOKLOC{} LOC & $\SCCTESTS{}/\SCCTESTS{} + \SCEXAMPLES{}/\SCEXAMPLES{}^{\text{a}}$ \\
\code{uvm-systemc} (\code{\UVMCOMMIT}) & 1800.2-2020 UVM/RAL & 78 br. & $\UVMCTESTS{}/\UVMCTESTS{}$ (100\%) \\
\code{crave} (\code{\CRAVECOMMIT}) & 1800 Cl.~18 CRV & 39 br. & $\CRAVECTESTS{}/\CRAVECTESTS{}$ (100\%) \\
\code{fc4sc} (\code{\FCFOURCOMMIT}) & Cl.~16/19 SVA + UCIS & 37 br. & $\FCFOURCTESTS{}/\FCFOURCTESTS{}^{\text{b}}$ (100\%) \\
\code{systemc-lp} (\code{\SCLPCOMMIT}) & 1801 UPF~3.1 runtime & \SCLPSRCLOC{} src LOC & $\SCLPCTESTS{}/\SCLPCTESTS{}$ (\SCLPTIME{})$^{\text{a,b}}$ \\
\code{svcc} (\code{\SVCCCOMMIT}) & SV$\leftrightarrow$SC + UPF miters & +\SVCCUPFADD{} rs LOC & 5/5 UPF; 45/50 SEC$^{\text{c}}$ \\
\code{InFormal} (\code{\INFCOMMIT}) & Direct LP + SVA FPV & \INFCRATES{} crates & Paths A \& B \\
\code{SystemC-Ex.} (\code{\EXCOMMIT}) & APB+TB (\APBSUBSYSLOC{} LOC; 234 RTL) & 5 LP bins & $\SMOKEPASS{}/\SMOKEPASS{}$ smoke (100\%) \\
\bottomrule
\end{tabularx}
\vspace{2pt}
\begin{flushleft}
\scriptsize $^{\text{a}}$Clean under Release and \code{ASan+UBSan}; \code{systemc-lp} totals \SCLPTOTALLOC{} LOC with tests and CMake; \code{br.} counts merged parity branches. $^{\text{b}}$All emitted UCIS~1.0 XML databases across \code{fc4sc}, \code{systemc-lp}, and \code{apb\_subsys} pass \code{xmllint {-}{-}schema UCIS.xsd}. $^{\text{c}}$45 inductively proved (0 failures); 5 multi-clock/array-port cases rejected fail-closed.
\end{flushleft}
\end{table}

\begin{table}[t]
\caption{2M-Cycle \code{apb\_pmu\_timer\_subsys} Runtime Overhead vs.\ Stock SystemC~3.0.3 (Apple M5~Max, 9 Interleaved Runs)}
\label{tab:rq1_perf}
\centering
\scriptsize
\setlength{\tabcolsep}{3.0pt}
\begin{tabularx}{\columnwidth}{@{}>{\raggedright\arraybackslash}Xrrr@{}}
\toprule
\textbf{Configuration} & \textbf{Median (IQR)} & \textbf{Speed} & \textbf{vs.\ Stock} \\
\midrule
Stock SystemC~3.0.3 (unhooked) & 159.6\,ms (3.5) & 12.53\,Mcyc/s & baseline \\
Mode 1b: CLI LP-OFF (no \code{{-}{-}upf}) & 165.8\,ms (3.0) & 12.07\,Mcyc/s & $+3.8\%$ \\
Mode 2: \code{SystemC::lp\_stub} & 166.6\,ms (2.1) & 12.00\,Mcyc/s & $+4.4\%$ \\
Mode 3: \code{-DSYSTEMC\_LP\_DISABLE} & 165.6\,ms (2.6) & 12.07\,Mcyc/s & $+3.8\%$ \\
Mode 1a: LP-ON (100\% active) & 864.4\,ms (18.5) & 2.31\,Mcyc/s & $5.42\times$ \\
Mode 1a: LP-ON (50\% gated \code{OFF}) & 807.7\,ms (6.9) & 2.48\,Mcyc/s & $5.06\times$ \\
\bottomrule
\end{tabularx}
\end{table}

To measure kernel hook overhead, we benchmarked $2 \times 10^6$ clock cycles of \code{apb\_pmu\_timer\_subsys} across nine interleaved rounds after two warm-up runs against an unhooked build of upstream SystemC~3.0.3 (Table~\ref{tab:rq1_perf}). Because \code{apb\_pmu\_timer\_subsys} contains only two small modules, kernel scheduler dispatch dominates cycle execution and yields a conservative upper bound on hook overhead. All three deactivation modes sustain $12.0\,$Mcyc/s and remain within $3.8\%$--$4.4\%$ of unhooked SystemC~3.0.3; their medians sit within the $2$--$3\,$ms interquartile spread because all three execute the same inline \code{m\_power\_ext\_p} and \code{is\_power\_gated()} checks inside \code{libsystemc}. Enabling full UPF~3.1 execution (Mode~1a, with supply-switch solving, strategy evaluation, and per-timestep PST legality and level-shifter checks) runs at $2.31\,$Mcyc/s; gating \code{PD\_TIM} off for a 50\% sleep duty cycle trims runtime by $6.6\%$ ($2.48\,$Mcyc/s) by suppressing \code{u\_timer::seq\_proc} dispatch while asleep.

\subsection{Dynamic Fault-Injection and Sign-Off Suite}
\label{sec:case_sc}

\begin{table*}[t]
\caption{Low-Power Hardware, Firmware, SRAM, and Testbench Defect Classes Verified Dynamically in \code{systemc-lp}}
\label{tab:bugs}
\centering
\footnotesize
\setlength{\tabcolsep}{5.0pt}
\begin{tabularx}{\textwidth}{@{}>{\raggedright\arraybackslash}p{3.6cm}>{\raggedright\arraybackslash}p{5.5cm}>{\raggedright\arraybackslash}X@{}}
\toprule
\textbf{Defect Class} & \textbf{Injected RTL / PMU / UPF Fault} & \textbf{Detection Mechanism in \code{systemc-lp} and Header-Only Adapters} \\
\midrule
1. Premature power cut (\code{LP\_BUG=1}) & PMU drops \code{tim\_pwr\_en\_q} before asserting \code{tim\_iso\_en\_q} on \code{PD\_TIM} outputs. & Raises \code{ISO\_VIOLATION}, fires Preponed isolation SVA, and scrambles unisolated \code{PRDATA}/\code{IRQ} into \code{PD\_AON}. \\
\addlinespace[1.5pt]
2. Late retention save (\code{LP\_BUG=2}) & PMU pulses \code{tim\_ret\_save\_q} after switched rail \code{VDD\_TIM} leaves \code{FULL\_ON}. & Raises \code{RET\_HAZARD} and invalidates shadow latch; post-wakeup \code{tim\_load->read()} catches corrupted \code{0x0024} reload CSR. \\
\addlinespace[1.5pt]
3. Uninitialized cold wake (\code{LP\_BUG=3}) & Firmware reads volatile \code{scratch\_q} (\code{0xCAFE}) after cold wake before re-writing. & Two-state scrambling corrupts \code{scratch\_q}; \code{bridges/uvm.h} \code{post\_read} raises \code{RAL\_CORRUPT} alongside CSR readback mismatch. \\
\addlinespace[1.5pt]
4. Multi-strategy \& glitch ISO violation & \code{iso\_en} glitches low during \code{CORRUPT}, or one of multiple domain strategies stays low. & Continuous \code{CORRUPT}-window monitor and per-domain \code{dom\_iso\_conj} conjunction flag \code{ISO\_VIOLATION} while preserving per-driver \code{sc\_signal\_rv} clamps. \\
\addlinespace[1.5pt]
5. Retention brownout, hold, \& reset clobber & Standby rail droops below \code{add\_port\_state} floor, \code{-retention\_condition} drops, or reset clobbers restore. & Standby voltage-floor and hold monitors scramble shadow latches (\code{RET\_HAZARD}); post-restore reset check flags \code{RET\_CLOBBERED\_BY\_RESET}. \\
\addlinespace[1.5pt]
6. Dual-rail SRAM write-disturb & Unisolated write enable mutates \code{VlUnpacked}/C++ array while \code{VDD\_PERIPH == OFF} and \code{VDDM == FULL\_ON}. & \code{sclp\_memory.h} compares periphery-collapse array against pre-sleep snapshot, scrambles disturbed words, and flags \code{RET\_HAZARD}. \\
\addlinespace[1.5pt]
7. Stale TLM DMI \& illegal PST arc & Initiator reuses cached DMI pointer across collapse, or PMU jumps across forbidden PST state pair. & \code{sclp\_tlm\_domain\_guard} calls \code{invalidate\_direct\_mem\_ptr()}; \code{add\_pst\_transition} check flags \code{PST\_ILLEGAL} while \code{SCLP\_ASSERT\_DOMAIN\_PROPERTY} gates off-domain SVA. \\
\addlinespace[1.5pt]
8. Overlapping switch inrush ramp & Multiple switches on a shared parent rail overlap inside \code{-ack\_delay} turn-on windows (\code{set\_inrush\_limit}). & Supply solver tracks concurrent \code{pending\_rise} intervals, flagging inrush violations while passing daisy-chained \code{ack $\rightarrow$ en} startups. \\
\addlinespace[1.5pt]
9. Time-zero UPF/PST mismatch & Unvisited DVFS row lacks \code{set\_level\_shifter}, ISO/RET uses switched self rail, or switches share unstaggered control. & \code{{-}{-}lp-lint} inspects all elaborated crossings and PST rows at $t = 0$, reporting structural errors before cycle~0. \\
\bottomrule
\end{tabularx}
\vspace{-4pt}
\end{table*}

In \code{SystemC-Examples}, a single \code{top.cpp} testbench for \code{apb\_pmu\_timer\_subsys} compiles five binaries covering three positive CTests (baseline functional execution, \code{SystemC::lp\_stub} link swapping, and full \code{{-}{-}upf} closure across all three PST states and all six legal state transitions) plus three fault-injected binaries (\code{-DLP\_BUG=1..3}). Combined with \code{systemc-lp}'s \SCLPCTESTS{} unit, mixed-language \code{Verilator {-}{-}sc} + \code{package UPF} DPI, dual-rail SRAM, TLM-2.0 DMI, staggered inrush, and static-lint regressions, Table~\ref{tab:bugs} catalogs the nine defect classes verified dynamically. Every injected fault produces a deterministic diagnostic and test failure without false alarms on legal power sequences.

\subsection{Word-Level Formal Proofs and Narrow-Cone Mutations}
\label{sec:formal_eval}

\begin{table}[t]
\caption{\code{svcc prove-upf} and \code{InFormal} Verification Results on \code{apb\_pmu\_timer\_subsys}}
\label{tab:formal_timings}
\centering
\scriptsize
\setlength{\tabcolsep}{2.8pt}
\begin{tabularx}{\columnwidth}{@{}l>{\raggedright\arraybackslash}Xrr@{}}
\toprule
\textbf{Engine \& Check} & \textbf{Target / Scope} & \textbf{Verdict} & \textbf{Time} \\
\midrule
\code{svcc upf-static} & Structural UPF + port paths & 0 findings & $<1.0\,$ms \\
\code{svcc cpp-ub} & \code{svcc\_\_ub == 0} (6 steps) & UNSAT & 13.35\,ms \\
\code{svcc iso-2copy} & \code{PD\_TIM} $\rightarrow$ 4 top outputs & UNSAT & 8.84\,ms \\
\code{svcc retention} & \code{u\_timer.cfg\_q} (8-bit, 4-step) & UNSAT & 9.48\,ms \\
\code{svcc retention} & \code{u\_timer.load\_q} (16-bit, 4-step) & UNSAT & 9.10\,ms \\
\code{svcc proto-bmc} & Unconstrained APB / read-only & SAT / UNSAT & 34.84\,ms \\
\midrule
Path A \code{load\_upf} & PST (22 CNF vars, 47 clauses) & SAT (valid) & 1.82\,ms \\
Path A \code{ISO\_MISSING} & \code{u\_timer/svcc\_\_ub} dead cone & Waived (COI) & 0.00\,ms \\
Path A \code{check\_lp} & \code{PREADY}, \code{PSLVERR}, \code{svcc\_\_ub} & Proven (3/5 outs) & 42--70\,ms \\
Path A \code{LP\_CORRUPT} & \code{PRDATA}, \code{IRQ} (unconstrained APB) & CEX (2/5, $d = 2$) & 46--48\,ms \\
\midrule
Path B Covers & Off-ISO, PST$\times 3$, full RET ($d=10$) & Covered (5/5) & 12--19\,ms \\
Path B RET SVA & \code{cfg\_q}, \code{load\_q} ($k=1$ induction) & Proven (6/6) & 22--28\,ms \\
Path B PMU SVA & 2QBF-validated FW CSR assume & Proven (4/4) & 20--25\,ms \\
Path B PST SVA & \code{SUBSYS\_PST} (\code{\{LV\_0V75, OFF\}} gap) & CEX (1/1, $d = 2$) & 30\,ms \\
\bottomrule
\end{tabularx}
\end{table}

Passing the unmodified \code{apb\_pmu\_timer\_subsys.h} header and port-scoped \code{apb\_subsys.upf} script through \code{svcc prove-upf} and \code{InFormal} Paths~A and~B finishes every check within 70\,ms (Table~\ref{tab:formal_timings}). In \code{svcc prove-upf}, Z3 proves C++17 undefined-behavior freedom (\code{cpp-ub}), two-copy isolation non-interference across all four top-level outputs (\code{iso-2copy}, 8.84\,ms), and sleep--wake retention recovery from unconstrained initial states on both \code{cfg\_q} and \code{load\_q}. Four controlled mutations in \code{upf\_formal.rs} test narrow-cone sensitivity. Removing \code{u\_timer/tim\_rdata} from \code{ISO\_TIM} exposes combinational corruption on \code{PRDATA} during the APB setup phase (\code{PENABLE == 0}). Gating an unisolated 32-bit bus through \code{fatal\_irq = (tag\_w == 32'hDEAD\_BEEF)} yields a single-query \code{iso-2copy} counterexample despite its $2^{-32}$ random hit rate in simulation, and proves UNSAT once isolated. Pulsing \code{save} after primary rail collapse (\code{LP\_BUG=2}) refutes \code{retention}, while introducing an unshifted cross-voltage PST row or self-supplied isolation strategy flags \code{LS\_MISSING} and \code{ISO\_WHILE\_OFF}.

\code{InFormal} Paths~A and~B confirm and extend those results. Path~A checks PST satisfiability in \code{InFormal}'s CaDiCaL-backed SAT solver (1.82\,ms), prunes the unreachable \code{u\_timer/svcc\_\_ub} crossing via cone-of-influence analysis without manual waivers, and proves three of five outputs (\code{PREADY}, \code{PSLVERR}, \code{svcc\_\_ub}) immune to off-domain corruption; on \code{PRDATA} and \code{IRQ}, unconstrained APB writes produce depth-2 unisolated-shutdown counterexamples (\code{PWDATA=0x00}), and even under isolation-before-off sequencing volatile \code{scratch\_q} and \code{count\_q} leak across wakeup unless \code{tim\_por\_n\_q = 0} resets \code{count\_q} and firmware re-writes \code{scratch\_q}. On the 16-property UPF-lowered SystemVerilog netlist, Path~B hits all five witness covers (up to a 10-step save--collapse--restore cycle), proves all six retention shadow assertions by 1-step induction ($k = 1$), flags the missing \code{\{LV\_0V75, OFF\}} row in \code{SUBSYS\_PST} at depth~2, and validates the APB firmware write-ordering contract via 2QBF \code{proposal\_validate} before proving all four PMU sequencing assertions.

\begin{table*}[t]
\caption{Engine and Subsystem Defects Uncovered by Cross-Checking \code{systemc-lp}, \code{svcc}, and \code{InFormal} on Shared Design Files}
\label{tab:cross_bugs}
\centering
\scriptsize
\setlength{\tabcolsep}{4.2pt}
\begin{tabularx}{\textwidth}{@{}ll>{\raggedright\arraybackslash}p{4.8cm}>{\raggedright\arraybackslash}X@{}}
\toprule
\textbf{ID} & \textbf{Exposed By} & \textbf{Root Cause in Engine or Subsystem} & \textbf{Why Initial 2-State Sim Missed It \& Verified Resolution} \\
\midrule
\code{SCLP-BUG-01} & \code{InFormal check\_lp} (69\,ms) & Elaborator matched \code{channel->name()} (\code{dut.tim\_irq}) only, ignoring bound child \code{sc\_port\_base} paths (\code{u\_timer/tim\_irq}). & Early tests omitted \code{-elements} or named parent wires. Fixed via \code{walk\_ports()} child-port binding. \\
\addlinespace[0.5pt]
\code{SCLP-BUG-02} & \code{svcc upf\_lower} audit + CTest & Strategy loop overwrote \code{dom->iso\_asserted = active}, letting a second strategy mask an unasserted strategy on the same domain. & Prior unit tests attached one strategy per domain. Fixed via conjunctive \code{dom\_iso\_conj} while keeping \code{sc\_signal\_rv} clamps per-driver. \\
\addlinespace[0.5pt]
\code{SCLP-BUG-03} & \code{svcc upf-static} comparison & Unmatched UPF \code{-elements}, \code{-control\_port}, and \code{-save/-restore\_signal} paths failed open without diagnostics. & Positive tests used valid paths; fault tests injected timing errors. Fixed by raising \code{ELABORATION} diagnostics on unmatched targets. \\
\midrule
\code{SUBSYS-01} & \code{InFormal check\_lp} (69 / 48\,ms) & Initial \code{apb\_subsys.upf} listed parent wires \code{\{tim\_irq tim\_rdata\}} instead of IEEE~1801-2018 \S6.45 child ports. & \code{systemc-lp} matched parent wires only. Updated \code{ISO\_TIM} to child port paths (0 unwaived violations). \\
\addlinespace[0.5pt]
\code{SUBSYS-02} & \code{svcc iso-2copy} (8.84\,ms) & \code{fabric\_mux()} forced \code{PRDATA = 0} only when \code{PSEL \&\& PENABLE}; setup cycle (\code{PENABLE == 0}) passed raw \code{tim\_rdata}. & APB bus monitors sample \code{PRDATA} only when \code{PENABLE \&\& PREADY}. Caught in one query when \code{tim\_rdata} is unisolated. \\
\addlinespace[0.5pt]
\code{SUBSYS-03} & \code{InFormal formal\_lint} ($<5\,$ms) & \code{SV-FPV-002}: retained \code{u\_timer.cfg\_q} omits \code{por\_n} reset so warm sleep keeps config, leaving \code{cfg\_q[1:0]} uninitialized at cold boot. & C++ \code{sc\_uint<8>} zero-initializes at elaboration, masking cold power-on randomness on \code{en} and \code{irq\_en}. Boot sequence writes \code{TIM\_CFG} explicitly prior to interrupt enable. \\
\addlinespace[0.5pt]
\code{SUBSYS-04} & \code{InFormal check\_lp} (46--48\,ms) & Unconstrained \code{PWDATA=0x00} leaks unisolated \code{PRDATA}/\code{IRQ} ($d = 2$); under ISO, volatile \code{scratch\_q}/\code{count\_q} leak on wake without \code{tim\_por\_n\_q = 0} and CSR re-init. & Directed wakeup test sequenced ISO, pulsed \code{tim\_por\_n\_q = 0}, and reloaded CSRs. \code{lp\_diagnose} extracted the 42-signal causal slice. \\
\addlinespace[0.5pt]
\code{SUBSYS-05} & \code{svcc} (34.8\,ms) \& Path~B & Single APB write \code{PWDATA=0x00} cuts power without ISO, and \code{\{LV\_0V75, OFF\}} was missing from \code{SUBSYS\_PST}. & Directed UVM sequence issued legal step-by-step PMU writes. Formalized primary-input CSR contract via \code{proposal\_validate} (4/4 proven). \\
\addlinespace[0.5pt]
\code{SUBSYS-06} & \code{InFormal} (0.74\,ms) \& \code{svcc} & \code{LS\_TIM} flagged \code{LS\_REDUNDANT} ($\times 10$) under symmetric DVFS PST; asymmetric \code{\{0.9V, 0.75V\}} without LS flags \code{LS\_MISSING}. & Dynamic test checked illegal voltage drops, not redundant shifters under symmetric DVFS rows. \\
\bottomrule
\end{tabularx}
\vspace{-4pt}
\end{table*}

\subsection{Bidirectional Bug Discovery Across Independent Tools}
\label{sec:bug_discovery}

A direct test of macro-free source reuse came when we ran \code{svcc} and \code{InFormal} on files that had already passed our initial \code{systemc-lp} regressions. Cross-checking the same C++ RTL and \code{.upf} script uncovered three bugs inside \code{systemc-lp} (\code{SCLP-BUG-01..03}) and six subsystem RTL, UPF, and firmware issues (\code{SUBSYS-01..06}), listed in Table~\ref{tab:cross_bugs}.

Running \code{InFormal check\_lp} on our original \code{apb\_subsys.upf} (which named parent signals \code{\{tim\_irq tim\_rdata\}}) reported \code{ISO\_MISSING} (\code{SUBSYS-01}). Rewriting \code{-elements} to IEEE~1801 child port paths (\code{\{u\_timer/tim\_irq u\_timer/tim\_rdata\}}) cleared both formal tools, but revealed that \code{systemc-lp} matched only parent \code{sc\_signal} names (\code{SCLP-BUG-01}) and ignored unresolved UPF targets without warning (\code{SCLP-BUG-03}). Comparing our runtime evaluator against \code{svcc}'s conjunctive isolation equation ($\bigwedge_i \text{act}_i$) then exposed a last-writer-wins overwrite of \code{dom->iso\_asserted} when two strategies shared one domain (\code{SCLP-BUG-02}). When our first patch for \code{SCLP-BUG-02} clamped resolved buses across the entire channel, \code{systemc-lp}'s existing tri-state \code{sc\_signal\_rv} regression caught the resulting over-clamping of an always-on peer driver.

In the reverse direction, formal analysis exposed five design and firmware corner cases that passing two-state tests had missed: setup-phase \code{PRDATA} leakage while \code{PENABLE == 0} (\code{SUBSYS-02}), elaboration-time C++ zero-initialization hiding the resetless \code{cfg\_q[1:0]} control bits at cold boot (\code{SUBSYS-03}), post-wakeup corruption paths from \code{scratch\_q} and \code{count\_q} when firmware wakes \code{PD\_TIM} without pulsing \code{tim\_por\_n\_q = 0} and re-initializing volatile CSRs (\code{SUBSYS-04}), missing hardware CSR interlocks and omitted PST rows (\code{SUBSYS-05}), and redundant level shifters under synchronous DVFS (\code{SUBSYS-06}).

\subsection{Threats to Validity and Scope Boundaries}
\label{sec:limitations}

Two groups of scope boundaries frame our findings. First, because IEEE~1801 files execute as Tcl programs, \code{systemc-lp}, \code{svcc}, and \code{InFormal} verify the elaborated power model for a chosen design configuration rather than unexecuted Tcl branches. Within UPF~3.1, \code{systemc-lp} implements 28 RTL commands (plus \code{add\_pst\_transition} and \code{set\_inrush\_limit}), while \code{svcc-ir::upf} lowers a 22-command structural subset (with \code{set} and \code{foreach} expansion and fail-closed rejection of procedural \code{if}/\code{proc}/\code{expr} constructs); Liberty cell-level power tables and UPF~4.0 analog extensions remain outside scope.

Second, \code{systemc-lp} models discrete rail states, microvolt retention floors, and digital \code{-ack\_delay} switch ramp windows rather than continuous IR drop, LDO transient droop, or SPICE-level retention metastability, which require co-simulation with SystemC-AMS macromodels. On the formal side, Btor2 lowering assumes a single root clock and edge-sampled resets, so multi-clock wakeup handshakes and large dual-rail SRAM arrays are verified in dynamic \code{systemc-lp} runs; extending Table~\ref{tab:rq1_perf}'s throughput study to multi-core SoCs booting full operating systems is natural follow-on work.

%% file: sections/related_work.tex
\section{Related Work}
\label{sec:related}

\begin{table*}[t]
\caption{Comparison of Low-Power SystemC, RTL Simulation, and Formal Verification Approaches}
\label{tab:related}
\centering
\scriptsize
\setlength{\tabcolsep}{3.6pt}
\begin{tabularx}{\textwidth}{@{}>{\raggedright\arraybackslash}Xcccccc@{}}
\toprule
\textbf{Capability} & \textbf{PKtool}~\cite{vece2009pktool,vece2011pktool} & \textbf{PwARCH / TLM}~\cite{lebreton2008pwarch,mbarek2012upftlm,mischkalla2013patmos} & \textbf{SCPower}~\cite{gagarski2016scpower} & \textbf{Vendor PA-Sim}~\cite{bembaron2009dvcon} & \textbf{POWER-TRUCTOR}~\cite{hazra2013powertructor} & \textbf{This Work} \\
\midrule
Primary abstraction & SC ESL/RTL & SC TLM-2.0 & Macro SC & 4-state SV & Formal SV & SC/SV RTL, Btor2 \\
Unmodified RTL (0 LP macros) & Partial & No (TLM hooks) & No (C++ macros) & Yes & Yes & Yes \\
IEEE~1801 Tcl role & None & TLM subset & Generates UPF & Full 1801 Tcl & Static extract & 28-cmd Tcl (+2; 22 formal) \\
Cycle-level corrupt \& retention & No (energy) & No (socket) & Yes (macros) & Yes (\code{'X}) & Symbolic & Yes (\code{'X}/\code{'Z} + jitter) \\
Process gating while \code{OFF} & No & Decoupled & No & Simulator-specific & N/A & \code{sc\_process\_b} gate \\
Functional turnoff modes & None & None & Recompile & Elab.\ flags & N/A & CLI, link stub, elide \\
UVM RAL + CDV + UCIS~1.0 & No & No & No & SV \code{uvm\_pkg} & No & C++17 header adapters \\
Shared-source SEC \& X-prop & No & No & No & Separate tools & PMU SEC & \code{svcc} + \code{InFormal} \\
\bottomrule
\end{tabularx}
\end{table*}

Table~\ref{tab:related} compares \code{systemc-lp} and its formal verification flow against five lines of prior low-power verification research.

\textbf{SystemC energy estimation, TLM-2.0 prototypes, and macro libraries.} Early SystemC power frameworks such as \code{PKtool}~\cite{vece2009pktool,vece2011pktool} attach signal-activity observers and energy macromodels to \code{sc\_module} instances; because estimation hooks never corrupt registers, clamp boundary ports, or gate processes, functional bugs caused by reading volatile state after wake pass undetected. Transaction-level approaches including Lebreton and Vivet~\cite{lebreton2008pwarch}, Mbarek et al.'s \code{PwARCH}~\cite{mbarek2012upftlm}, and Mischkalla and Mueller~\cite{mischkalla2013patmos,mischkalla2014patmos} lift UPF domains and PSTs onto loosely-timed TLM-2.0 sockets so firmware teams can validate coarse power transitions early. While \code{systemc-lp} also guards TLM-2.0 sockets and DMI pointers via \code{bridges/tlm.h}, socket-level models operate above clock edges and pins and therefore cannot catch single-cycle shutdown races (\code{LP\_BUG=1}), late retention saves (\code{LP\_BUG=2}), or APB setup-phase multiplexer leaks (\code{SUBSYS-02}). Closer to cycle-accurate RTL, Gagarski et al.'s \code{SCPower}~\cite{gagarski2016scpower} instruments SystemC code with custom C++ power macros and generates UPF from those annotations; embedding macros inside RTL headers, however, breaks IEEE~1666.1 synthesis and formal compilers (\code{svcc}) and prevents executing pre-existing external \code{.upf} scripts on unmodified IP.

\textbf{Four-state power-aware RTL simulation and X-propagation.} Commercial Verilog simulators elaborate external \code{.upf} files against unmodified RTL~\cite{bembaron2009dvcon} using four-state logic (\code{0}, \code{1}, \code{X}, \code{Z}). As Turpin~\cite{turpin2003xbugs} showed, standard Verilog ternary semantics still suffer from X-optimism when \code{if (cond)} steers an \code{'X} condition deterministically into the \code{else} branch, and from X-pessimism when reconvergent logic such as \code{a \textasciicircum{} a} fails to cancel unknown bits. While \code{systemc-lp} drives \code{SC\_LOGIC\_X} and \code{SC\_LOGIC\_Z} on four-state SystemC signals, two-state C++17 execution in \code{sc\_uint} datapaths and \code{Verilator {-}{-}sc} omits \code{'X} storage for simulation speed, motivating multi-cycle \code{CORRUPT\_JITTER} scrambling for sticky single-bit paths alongside SAT/SMT reasoning for wide datapath decoders.

\textbf{Formal UPF verification and self-composition.} Hazra et al.~\cite{hazra2010dac,hazra2013powertructor} (\code{POWER-TRUCTOR}) and Dasgupta et al.~\cite{dasgupta2014pmu} extract properties from UPF intent to verify PMU controllers and hardware/software power sequencing formally, while Hyun and Kim~\cite{hyun2019srpg} optimize multi-bit state-retention register allocation and wakeup latency during power-gated synthesis. For unknown-value propagation, Kochte et al.~\cite{kochte2012satx} replace lossy ternary simulation with SAT reasoning, and Yang et al.~\cite{yang2018lazyself} and Tan et al.~\cite{tan2026secic3} (\code{SecIC3}) verify non-interference on two-copy self-composition miters using IC3, complemented by word-level vector preservation techniques such as Guimar\~{a}es et al.~\cite{guimaraes2026vectorization}. Our work connects dynamic C++17 UPF simulation and word-level self-composition/inductive proofs over the exact same unmodified SystemC/SystemVerilog RTL and \code{.upf} files; mismatches in either tool then expose defects in the other (\code{SCLP-BUG-01..03} and \code{SUBSYS-01..06}).

%% file: sections/conclusion.tex
\section{Conclusion}
\label{sec:conclusion}

Keeping synthesizable RTL free of low-power macros allows one design and one external \code{.upf} specification to drive both fast C++17 simulation and word-level formal verification. By adding a UPF-agnostic, \code{+\SCHOOKDIFF{}}-line policy interface to SystemC~3.0.3, \code{systemc-lp} executes 28 IEEE~1801 UPF~3.1 RTL commands (plus \code{add\_pst\_transition} and \code{set\_inrush\_limit}), time-zero static LP lint, orthogonal \code{sc\_process\_b} gating, port-scoped isolation clamping, retention brownout and dual-rail SRAM write-disturb checks, and both four-state \code{'X}/\code{'Z} and two-state \code{CORRUPT\_JITTER} corruption across SystemC and \code{Verilator {-}{-}sc} models, while staying within \LPOFFDELTA{} of unhooked SystemC when UPF is disabled. Header-only adapters synchronize UVM-SystemC register mirrors, TLM-2.0 DMI pointers, FC4SC off-domain assertions and UCIS~1.0 coverage, and CRAVE stimulus without modifying upstream Accellera libraries for low-power support. Feeding the identical RTL and \code{.upf} file into \code{svcc} and \code{InFormal} discharges word-level two-copy miters and inductive SVA contracts within 70\,ms, uncovering three implementation bugs in \code{systemc-lp} alongside six RTL, UPF, and firmware-sequencing defects missed by directed simulation.

\textbf{Availability.} All source repositories, reference designs, CMake helpers, and reproduction scripts evaluated in Table~\ref{tab:stack} are publicly available at \url{https://github.com/mayurkubavat} (\code{systemc-lp}, \code{oss-workbench}, the hooked Accellera forks, \code{SystemC-Examples}, \code{svcc}~\cite{kubavat2026svcc}, and \code{InFormal}~\cite{kubavat2026informal}).

\textbf{Use of AI.} Google Antigravity assisted with parts of the implementation, test scripts, and manuscript editing under the author's direction and review. The author designed the system architecture and formal encodings, ran and verified every measurement, proof, and reference reported here, and takes full responsibility for this work.

%% file: refs.bib
@manual{ieee1666_2023,
  title        = {{IEEE} Standard for Standard {SystemC} Language Reference Manual},
  organization = {IEEE Std 1666-2023},
  year         = {2023}
}

@manual{ieee1666_1_2016,
  title        = {{IEEE} Standard for Standard {SystemC} Language Reference Manual: Synthesizable Subset},
  organization = {IEEE Std 1666.1-2016},
  year         = {2016}
}

@manual{ieee1800_2023,
  title        = {{IEEE} Standard for {SystemVerilog}---Unified Hardware Design, Specification, and Verification Language},
  organization = {IEEE Std 1800-2023},
  month        = feb,
  year         = {2024}
}

@manual{ieee1800_2_2020,
  title        = {{IEEE} Standard for Universal Verification Methodology Language Reference Manual},
  organization = {IEEE Std 1800.2-2020},
  year         = {2020}
}

@manual{ieee1801_2018,
  title        = {{IEEE} Standard for Design and Verification of Low-Power, Energy-Aware Electronic Systems},
  organization = {IEEE Std 1801-2018 (UPF 3.1)},
  year         = {2018}
}

@manual{ieee1801_2024,
  title        = {{IEEE} Standard for Design and Verification of Low-Power, Energy-Aware Electronic Systems},
  organization = {IEEE Std 1801-2024 (UPF 4.0)},
  year         = {2024}
}

@manual{accellera_ucis,
  title        = {Unified Coverage Interoperability Standard ({UCIS}) Version 1.0},
  organization = {Accellera Systems Initiative},
  year         = {2012}
}

@misc{snyder_verilator,
  author       = {W. Snyder},
  title        = {{Verilator}: Open-Source {SystemVerilog} Simulator and Lint System},
  howpublished = {\url{https://www.veripool.org/verilator/}},
  year         = {2024}
}

@inproceedings{vece2009pktool,
  author    = {G. B. Vece and M. Conti},
  title     = {Power Estimation in Embedded Systems within a {SystemC}-Based Design Context: The {PKtool} Environment},
  booktitle = {Proc. IEEE 7th Int. Workshop on Intelligent Solutions in Embedded Systems (WISES)},
  year      = {2009}
}

@incollection{vece2011pktool,
  author    = {G. B. Vece and M. Conti},
  title     = {Power Analysis of Embedded Systems: The {PKtool} Simulation Environment},
  booktitle = {Solutions on Embedded Systems},
  series    = {Lecture Notes in Electrical Engineering},
  volume    = {81},
  pages     = {301--314},
  publisher = {Springer},
  year      = {2011}
}

@inproceedings{lebreton2008pwarch,
  author    = {H. Lebreton and P. Vivet},
  title     = {Power Modeling in {SystemC} at Transaction Level, Application to a {DVFS} Architecture},
  booktitle = {Proc. IEEE Computer Society Annual Symp. on VLSI (ISVLSI)},
  pages     = {463--466},
  year      = {2008}
}

@article{mbarek2012upftlm,
  author  = {O. Mbarek and A. Pegatoquet and M. Auguin},
  title   = {Using Unified Power Format Standard Concepts for Power-Aware Design and Verification of Systems-on-Chip at Transaction Level},
  journal = {IET Circuits, Devices \& Systems},
  volume  = {6},
  number  = {5},
  pages   = {287--296},
  year    = {2012}
}

@inproceedings{mischkalla2013patmos,
  author    = {F. Mischkalla and W. Mueller},
  title     = {Efficient Power Intent Validation Using Loosely-Timed Simulation Models},
  booktitle = {Proc. 23rd Int. Workshop on Power and Timing Modeling, Optimization and Simulation (PATMOS)},
  year      = {2013}
}

@inproceedings{mischkalla2014patmos,
  author    = {F. Mischkalla and W. Mueller},
  title     = {Advanced {SoC} Virtual Prototyping for System-Level Power Planning and Validation},
  booktitle = {Proc. 24th Int. Workshop on Power and Timing Modeling, Optimization and Simulation (PATMOS)},
  year      = {2014}
}

@inproceedings{gagarski2016scpower,
  author    = {K. Gagarski and M. Petrov and M. J. Moiseev and I. V. Klotchkov},
  title     = {Power Specification, Simulation and Verification of {SystemC} Designs},
  booktitle = {Proc. IEEE East-West Design \& Test Symp. (EWDTS)},
  year      = {2016}
}

@inproceedings{bembaron2009dvcon,
  author    = {F. Bembaron and R. Mukherjee and S. Kakkar and A. Srivastava},
  title     = {Low Power Verification Methodology Using {UPF}},
  booktitle = {Proc. Design and Verification Conf. (DVCon)},
  year      = {2009}
}

@inproceedings{turpin2003xbugs,
  author    = {M. Turpin},
  title     = {The Dangers of Living with an {X} (Bugs Hidden in Your {Verilog})},
  booktitle = {Proc. Synopsys Users Group (SNUG)},
  year      = {2003}
}

@inproceedings{hazra2010dac,
  author    = {A. Hazra and S. Mitra and P. Dasgupta and A. Pal and D. Bagchi and K. Guha},
  title     = {Leveraging {UPF}-Extracted Assertions for Modeling and Formal Verification of Architectural Power Intent},
  booktitle = {Proc. 47th ACM/IEEE Design Automation Conf. (DAC)},
  pages     = {773--776},
  year      = {2010}
}

@article{hazra2013powertructor,
  author  = {A. Hazra and R. Mukherjee and P. Dasgupta and A. Pal and K. Harer and A. Banerjee and S. Mukherjee},
  title   = {{POWER-TRUCTOR}: An Integrated Tool Flow for Formal Verification and Coverage of Architectural Power Intent},
  journal = {IEEE Trans. Comput.-Aided Design Integr. Circuits Syst.},
  volume  = {32},
  number  = {11},
  pages   = {1801--1813},
  year    = {2013}
}

@article{dasgupta2014pmu,
  author  = {P. Dasgupta and M. K. Srivas and R. Mukherjee},
  title   = {Formal Hardware/Software Co-Verification of Embedded Power Controllers},
  journal = {IEEE Trans. Comput.-Aided Design Integr. Circuits Syst.},
  volume  = {33},
  number  = {12},
  pages   = {2025--2029},
  year    = {2014}
}

@inproceedings{hyun2019srpg,
  author    = {G.-H. Hyun and T. Kim},
  title     = {Allocation of State Retention Registers Boosting Practical Applicability to Power Gated Circuits},
  booktitle = {Proc. IEEE/ACM Int. Conf. on Computer-Aided Design (ICCAD)},
  year      = {2019}
}

@article{kochte2012satx,
  author  = {M. A. Kochte and M. Elm and H.-J. Wunderlich},
  title   = {Accurate {X}-Propagation for Test Applications by {SAT}-Based Reasoning},
  journal = {IEEE Trans. Comput.-Aided Design Integr. Circuits Syst.},
  volume  = {31},
  number  = {12},
  pages   = {1908--1919},
  month   = dec,
  year    = {2012}
}

@inproceedings{yang2018lazyself,
  author    = {W. Yang and Y. Vizel and P. Subramanyan and A. Gupta and S. Malik},
  title     = {Lazy Self-Composition for Security Verification},
  booktitle = {Proc. Int. Conf. on Computer Aided Verification (CAV)},
  series    = {LNCS},
  volume    = {10982},
  pages     = {136--156},
  year      = {2018}
}

@inproceedings{tan2026secic3,
  author    = {Q. Tan and A. Gaonkar and Y.-W. Fan and A. Gupta and S. Malik},
  title     = {{SecIC3}: Customizing {IC3} for Hardware Security Verification},
  booktitle = {Proc. Design, Automation \& Test in Europe (DATE)},
  year      = {2026}
}

@inproceedings{demoura2008z3,
  author    = {L. de Moura and N. Bj{\o}rner},
  title     = {{Z3}: An Efficient {SMT} Solver},
  booktitle = {Proc. Tools and Algorithms for the Construction and Analysis of Systems (TACAS)},
  series    = {LNCS},
  volume    = {4963},
  pages     = {337--340},
  year      = {2008}
}

@inproceedings{niemetz2018btor2,
  author    = {A. Niemetz and M. Preiner and C. Wolf and A. Biere},
  title     = {{Btor2}, {BtorMC} and {Boolector} 3.0},
  booktitle = {Proc. Int. Conf. on Computer Aided Verification (CAV)},
  series    = {LNCS},
  volume    = {10981},
  pages     = {587--595},
  year      = {2018}
}

@inproceedings{biere2024cadical,
  author    = {A. Biere and T. Faller and K. Fazekas and M. Fleury and N. Froleyks and F. Pollitt},
  title     = {{CaDiCaL} 2.0},
  booktitle = {Proc. Int. Conf. on Computer Aided Verification (CAV)},
  series    = {LNCS},
  volume    = {14681},
  pages     = {133--152},
  year      = {2024}
}

@inproceedings{yu2021certifaiger,
  author    = {E. Yu and A. Biere and K. Heljanko},
  title     = {Progress in Certifying Hardware Model Checking Results},
  booktitle = {Proc. Int. Conf. on Computer Aided Verification (CAV), Part II},
  series    = {LNCS},
  volume    = {12760},
  pages     = {363--386},
  year      = {2021}
}

@misc{guimaraes2026vectorization,
  author       = {M. F. O. Guimar{\~a}es and U. Rosa and I. Trudel and J. V. A. Vieira and A. A. Mafra and M. Crepalde and F. M. Q. Pereira},
  title        = {Vectorization of {Verilog} Designs and its Effects on Verification and Synthesis},
  howpublished = {arXiv:2603.17099},
  year         = {2026}
}

@misc{kubavat2026svcc,
  author       = {M. Kubavat},
  title        = {Translation Validation of {SystemVerilog}-to-{SystemC} {RTL} Conversion by Reading the Generated {C++}},
  howpublished = {arXiv:2611.31674},
  year         = {2026}
}

@misc{kubavat2026informal,
  author       = {M. Kubavat},
  title        = {{InFormal}: An Open-Source Formal Property Verifier for {SystemVerilog} with Checked Verdicts},
  howpublished = {Tech.\ Rep., Nov.\ 2026, doi: 10.5281/zenodo.24261270},
  year         = {2026}
}
